# Generating synthetic citation networks with communities

Łukasz Brzozowski [a,*], Marek Gagolewski [a,b], Grzegorz Siudem [c]

[a] Warsaw University of Technology, Faculty of Mathematics and Information Science, ul. Koszykowa 75, 00-662 Warsaw, Poland
[b] Systems Research Institute, Polish Academy of Sciences, ul. Newelska 6, 01-447 Warsaw, Poland
[c] Warsaw University of Technology, Faculty of Physics, ul. Koszykowa 75, 00-662 Warsaw, Poland

A B S T R A C T

Generating realistic synthetic citation, patent, or component dependency networks is essential for benchmarking community detection, graph visualisation, and network data mining algorithms. We present a systematic comparison of generators of directed graphs that are nearly acyclic and have a ground-truth community structure. We evaluate 12 methods across 7 real citation networks and 26 metrics. We propose the practice of reversing the directions of edges in static generators to break cycles and induce a citation-like flow, which markedly improves the performance of a degree-corrected Stochastic Block Model. Our methodological approach to evaluating community detection benchmarks distinguishes between endogenous and exogenous mesoscopic similarities, with the latter proving more important. This distinction reveals that high-parameter models gain much of their advantage by memorising planted community statistics, which limits their fidelity when the evaluation is blind to those labels. Finally, we introduce the Citation Seeder (CS) algorithm, an iterative generator grounded in the Price-Pareto model of citation networks, with interpretable parameters and $O(N + E)$ generation. CS achieves competitive results against the best-performing baselines while using up to four orders of magnitude fewer parameters, providing an interpretable account of a network's structure and, on a temporal hold-out, extrapolating its future in-degree distribution more faithfully than descriptive baselines.

## 1. Introduction

Citation and patent networks are amongst the most studied real-world networks. They are characterised by a structurally distinct topology: they are directed, nearly acyclic graphs (near-DAGs) that exhibit community structures representing scientific fields or patent classifiers. By a near-DAG we mean a directed graph that becomes acyclic once a small ratio of edges – the back-edges, typically 3–5% in citation networks – is removed. In an idealised publication process, citations point strictly from newer to older work, giving a Directed Acyclic Graph; in practice, revisions and near-simultaneous publications introduce a few back-edges. We use “nearly acyclic graph” and “near-DAG” interchangeably. Despite their prevalence [1], the synthetic generation of this specific graph family remains underserved. Existing community-aware generators produce undirected or densely cyclic graphs that lack the temporal citation flow structure (new papers cite older ones). This limits our ability to reliably benchmark community detection algorithms, test link-prediction models, and study network growth processes on controlled synthetic data.

More precisely, the scale-free nature of citation distributions has long been explained through the lens of cumulative advantage, whereby papers that have been cited many times are more likely to be cited again, resulting in a power-law-like tail [2,3]. Derek de Solla Price was among the first to formalise this mechanism in the context of scientific networks, proposing a growth model

* Corresponding author.
*Email addresses:* lukasz.brzozowski@pw.edu.pl (Ł. Brzozowski), marek.gagolewski@pw.edu.pl (M. Gagolewski), grzegorz.siudem@pw.edu.pl (G. Siudem).
*URLs:* https://www.gagolewski.com/ (M. Gagolewski), https://siudem.fizyka.pw.edu.pl/ (G. Siudem).

where new papers cite existing ones preferentially based on their in-degrees [4]. This concept, rooted in earlier ideas like Yule's and Simon's "rich-get-richer" processes, was later generalised by the Barabási–Albert model, which introduced continuous growth and linear preferential attachment to explain scale-free properties across networks [3,5]. However, pure preferential attachment exhibits a strong bias towards older nodes, preventing newer papers from gaining prominence. To address this, subsequent models incorporated intrinsic node *fitness* – an inherent quality or attractiveness of papers that modulates attachment probability, as in the Bianconi–Barabási fitness model, enabling high-fitness latecomers to outperform incumbents [6,7]. Additionally, *temporal attention decay* (or ageing) mechanisms were introduced, often modelled via exponential or log-normal functions that diminish a paper's citability over calendar time, better capturing the empirical rise-and-fall life cycle of scientific impact [8]. Cai et al. [9] likewise pair a generative model with an analytical account of what it produces, deriving the polarisation and robustness of journal impact factors semi-analytically from an agent-based model of the publication cycle; their target, however, is the temporal dynamics of journal-level citation indices, not the network's mesoscopic and topological structure evaluated against real graphs with community labels.

While growth models capture temporal dynamics, a parallel stream of research has focused on generating networks with planted modular structures. The Stochastic Block Model (SBM) [10] and its Degree-Corrected variant (DC-SBM) [11] serve as generative baselines for networks with latent group assignments. Other widely used approaches include the Configuration Model [12], which preserves exact degree sequences without community awareness, and the LFR Benchmark [13], designed to produce power-law degree and community-size distributions. Recent approaches also include adjusted modularity optimisation [14] and deep generative models: Graph Variational Autoencoders [15] and autoregressive graph models [16]. Alternative approaches have attempted to combine community-aware detection with growth dynamics [17]; while highly flexible, these methods often act as structural black boxes, lacking the analytical tractability and interpretability of pure random graph models, limiting their use solely to mechanistic tasks.

A notable gap exists between temporal citation-specific growth models and static community-aware graph generators. In a comprehensive survey of random graph modelling concepts, Drobyshevskiy and Turdakov [18] analysed numerous generators exploiting principles like preferential attachment and stochastic block models, revealing fundamental limitations in integrating community structure with growth dynamics akin to those of citation networks; similarly, Bonifati et al. [19] reviewed state-of-the-art graph generators (including SBM, LFR, and configuration models) and highlighted their inability to simultaneously capture realistic modularity, scale-free degrees, and directed acyclic topologies. Furthermore, classical community-aware models natively produce undirected graphs or densely cyclic directed graphs [20], deviating from the inherently acyclic temporal flow of citation networks. Consequently, they cannot be directly applied or compared to real citation data without undergoing a structural transformation.

Let us also note that the evaluation of community detection algorithms relies heavily on synthetic benchmarks with ground-truth labels. While the LFR benchmark is the *de facto* standard for undirected networks, no widely adopted equivalent exists for directed, nearly acyclic graphs. The existing propositions rely on deep learning models that cannot scale to real data [21].

Theoretical limits of community detectability have been rigorously established for static undirected models, notably the phase transitions identified by [22] for symmetric Stochastic Block Models, including the Kesten–Stigum (KS) threshold below which weak recovery (i.e., better-than-random guessing) becomes information-theoretically impossible [23,24]. These results distinguish scenarios where spectral methods or belief propagation succeed above the KS threshold, from those where no polynomial-time algorithm can detect communities reliably below it [25]. However, these theoretical bounds primarily apply to undirected, symmetric SBMs without temporal ordering or directionality, and extensions to directed or asymmetric cases reveal additional challenges, such as shifted thresholds in sparse directed SBMs [26,27].

Moreover, establishing such detectability limits is a fundamentally different task from providing generative benchmarks that mimic real-world structural complexity, especially for directed acyclic graphs like citation networks, where edges respect a temporal ordering and exhibit growth dynamics [1]. A benchmark for citation networks must not only plant ground-truth labels but also replicate the specific topological flow – preferential attachment within communities, ageing effects, and near-acyclic structure – that algorithms rely on to partition the graph accurately [21]. Static generators like LFR or degree-corrected SBMs produce undirected or cyclic directed graphs, failing to capture these traits without ad-hoc modifications [19]. While edge deletions have been proposed to remove cycles in, e.g., taxonomy networks [28], we are not aware of a systematic approach for cycle count reduction, rather than cycle deletion. Overall, adapting static generators to emulate citation-like flows is poorly understood. The practice of reversing the direction of edges to follow a heuristic topological ordering and reintroducing a fraction as back-edges has never been systematically evaluated for its effect on structural fidelity [29].

To address the foregoing issues, in this paper, we make three primary contributions (see Table 1 for an explanation of the notation).

1. **The CS generator.** We introduce the Citation Seeder (CS) algorithm, an iterative model grounded in the Price-Pareto model of citation networks [31]. This theoretical model distinguishes between accidental citations (allocated uniformly at random) and preferential citations (allocated proportionally to a vertex's in-degree, restricted to the node's community), yielding analytical, per-community Pareto Type II in-degree distributions. However, the 3DSI and Price-Pareto frameworks inherently produce analytical degree sequences rather than complete graph topologies. They predict citation accumulation but do not specify the edge list, DAG structure, or clustering patterns, which is a limitation that necessitates the development of a generative counterpart.

   To generate $k$ communities, CS uses $4k + 1$ analytically estimated parameters (where $k$ is the number of communities), generates in $O(N + E)$ time with $O(N \log N)$ parameter estimation, and natively outputs a near-DAG. Across 26 structural metrics on 7 real citation networks, CS achieves a competitive mean rank against the best-performing baseline, the Degree-Corrected Stochastic Block Model. CS achieves this result using up to four orders of magnitude fewer parameters and is not only parsimonious but also *explanatory*: its parameters are derived from a theory of network growth, with each parameter carrying a concrete bibliometric interpretation.

**Table 1**
Notation used throughout the paper. Symbols appearing only in the appendices are listed last.

| | | | |
|---|---|---|---|
| *Graph and model* | | | |
| $G = (V, E)$ | directed graph, vertices, edges | $N$, $N_i$ | no. of nodes; nodes in community $i$ |
| $k$ | no. of communities | $n_i$ | $N_i - 1$, edge-emitting nodes of community $i$ |
| $v$, $u$ | node being added; target node | $c(\cdot)$ | community-label map $V \to \{1, \dots, k\}$ |
| $d_{\text{in}}$, $d_{\text{out}}$ | in- and out-degree | $\hat{d}_{\text{out}}$ | mean out-degree of a dataset |
| $t$ | local time of a community | $\ell$ | node index within a community |
| *Model parameters* | | | |
| $p_i$ | growth probability | $m_i$ | expected out-degree |
| $\rho_i$ | preferentiality | $\sigma_i^2$ | out-degree variance |
| $\nu_i$ | effective preferentiality | $\langle a \rangle$ | global accidental citation rate |
| $\alpha_i$, $\lambda_i$ | Pareto type II shape and scale | $\Gamma(\cdot)$ | gamma function |
| *Generator* | | | |
| $\theta_i$ | negative-binomial shape | $\mu$ | Gamma-distributed Poisson rate |
| $n_{\text{acc}}$, $n_{\text{pref}}$ | accidental, preferential edge counts | $T$ | multiset of targets |
| $\text{Urn}_i$ | Pólya urn of community $i$ | $n_{\text{back}}$ | no. of injected back-edges |
| $r$ | back-edge ratio | $\beta$ | intra-community back-edge locality |
| $\uplus$ | multiset union | | |
| *Estimation* | | | |
| $\Psi_i^{\text{out}}$, $\Psi_i^{\text{in}}$ | total out-, in-degree of community $i$ | $\gamma_i$ | in-degree Gini index |
| $\hat{\cdot}$ | estimated quantity | $\epsilon$ | clamping tolerance, $10^{-3}$ |
| *Evaluation* | | | |
| $M$ | no. of compared methods (= 12) | $\bar{r}$ | mean rank |
| APE | absolute percentage error | $W$ | Wasserstein 1-distance |
| $L_1$ | $L_1$ distance | $\tau$ | triad type index |
| $s$ | category score, $(M + 1 - \bar{r})/M$ | $\Delta$ | rank change, endogenous excluded |
| $r_s$ | Spearman rank correlation | $\chi^2$ | Friedman statistic |
| $B$ | bootstrap resamples (= 10,000) | $Q$ | modularity |
| *Appendices* | | | |
| $\phi_i$ | over-dispersion ratio $\sigma_i^2/m_i$ | $S$ | target community size |
| $f$ | training fraction | $G_{\text{early}}$ | early (training) subgraph |
| $N_{\text{late}}$ | horizon size | $r_{\text{early}}$ | back-edge ratio of $G_{\text{early}}$ |

2. **Systematic analysis of cycle breaking.** We formalise the cycle breaking procedure and evaluate its impact on five classical generators across six structural categories. We demonstrate that cycle breaking is not uniformly beneficial; for example, it dramatically improves the degree structure of the Degree-Corrected Stochastic Block Model [11] while simultaneously degrading its mesoscopic exogenous fidelity.
3. **The endogenous/exogenous metric dichotomy.** We propose partitioning mesoscopic metrics into label-based (endogenous, such as modularity of ground truth partitions) and algorithm-behaviour-based (exogenous, such as modularity of algorithm-found partitions) categories. We argue that for the purpose of community-detection benchmarking, exogenous metrics are the primary concern. Decomposing the performance reveals that the models which encode ground-truth structure most directly gain much of their rank from endogenous planted statistics. When the evaluation is restricted to exogenous metrics, the $O(k)$ CS model overtakes the $O(N + k^2)$ DC-SBM. The effect is not one of parameter count as such: the $O(1)$ models (ER, LFR) rank last throughout, so parsimony is neither necessary nor sufficient for exogenous fidelity. What separates the models is *what* their parameters encode.

Overall, by introducing the CS generator, we aim to provide a unified framework for generating and researching citation-like networks. Unlike deep learning-based models that lack scalability and explainability [21], CS uses interpretable parameters to simulate field-specific growth rates, citation preferences, and inter-community interactions. Just like the methodological developments in the clustering of real-valued data [32–37], our tool allows researchers to evaluate community detection algorithms under specific, parametrised assumptions about how scientific fields grow, interact, merge, or split over time, facilitating studies of robustness to temporal biases and extrapolation of future degree growth [38].

## 2. The CS generator

In this section, we introduce the CS (Citation Seeder) algorithm, which is an iterative graph generator that produces directed near-DAGs with ground-truth community structure. It is inspired by the Price-Pareto model of citation networks with communities [31], extending that analytical framework from degree-sequence prediction to actual graph generation. This section presents the theoretical foundation, the engineering modifications that bridge theory and practice, the generation algorithm, and the parameter estimation procedure. A visualisation of an example network generated with the proposed algorithm is depicted in Fig. 1.

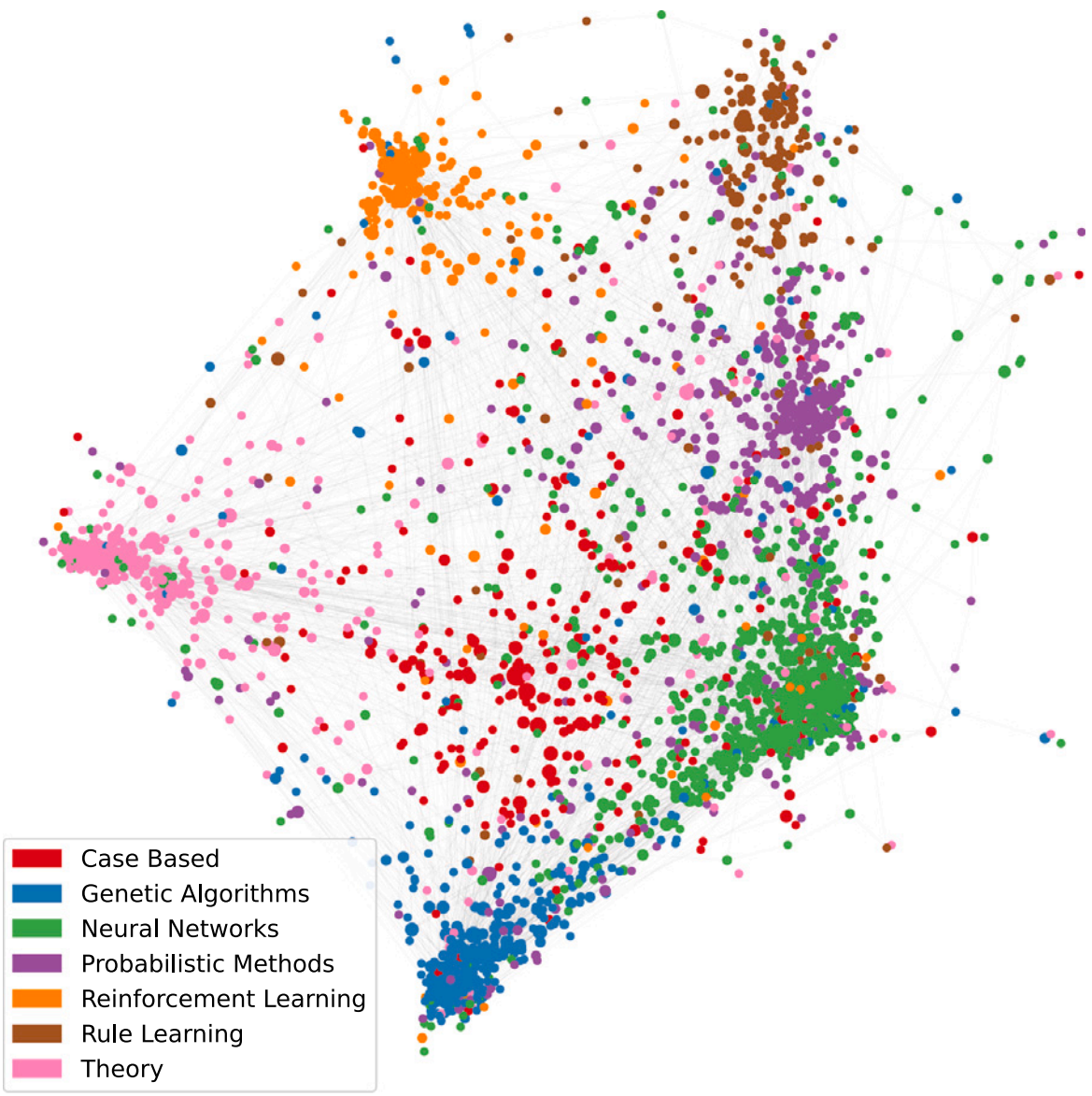

**Fig. 1.** An example network generated by our algorithm with parameters estimated from the Cora citation network [30]. The node colours correspond to the communities in the input network.

## 2.1. Theoretical foundation

Let us briefly summarise the Price-Pareto model with community structure [31]. The model assumes that a citation network grows one node at a time under the following assumptions:

(i) Each new node is assigned to community $i$ with probability $p_i > 0$, where $\sum_{i=1}^{k} p_i = 1$.
(ii) The new node creates $m_i$ directed edges to existing nodes.
(iii) A fraction $(1 - \rho_i)m_i$ of these edges is distributed *accidentally*: targets are chosen uniformly at random from the entire network, regardless of community. The remaining fraction $\rho_i m_i$ is *preferential*: targets are chosen with probability proportional to their current in-degree, restricted to the new node's own community.
(iv) Self-loops are disallowed. Time is local to each community: the "local time" $t$ of community $i$ equals its current size.

We assume discrete time. The model advances in global steps (one node is added per step across the whole network), but each community accumulates citations on its own clock: the local time $t$ of a community is its current node count, which increases only when a new node is assigned to *that* community. Because (1) describes in-degree growth within a single community $i$, $t$ throughout that expression refers to community $i$'s local time; we keep the unsubscripted $t$ to match the notation of the underlying model [31]. Since $p_i$ corresponds to the frequency with which community $i$ gains a new node, we have that the local time of community $i$ progresses approximately every $\frac{1}{p_i}$ global turns. Under these assumptions, we can derive a closed-form expression for the expected in-degree of node $\ell$ in community $i$ at local time $t$:

$$d_i^{(t)}(\ell) = \frac{\langle a \rangle}{\nu_i} \left[ \frac{\Gamma(\ell - \nu_i)}{\Gamma(\ell)} \cdot \frac{\Gamma(t)}{\Gamma(t - \nu_i)} - 1 \right], \tag{1}$$

where the *effective preferentiality* of community $i$ and the global accidental citation rate are defined, respectively, by:

$$\nu_i = \frac{\rho_i m_i}{\langle a \rangle + \rho_i m_i}, \qquad \langle a \rangle = \sum_{j=1}^{k} p_j m_j (1 - \rho_j). \tag{2}$$

The quantity $\nu_i \in (0, 1)$ can be interpreted as the fraction of each local time step during which preferential citations accumulate; it reduces to $\rho$ in the single-community case [39].

Two theoretical results are central to CS's design. First, as $t \to \infty$, the in-degree distribution in community $i$ converges to a *Pareto type II* (Lomax) distribution with shape $\alpha_i = 1/\nu_i$ and scale $\lambda_i = \langle a \rangle / \nu_i$. Second, the preferentiality parameter $\rho_i$, which is difficult to find, can be estimated under the model's assumptions from the network using, i.a., Gini indices of the in-degrees of all communities [40] (see Algorithm 2 for specific formulas). This connects in-degree inequality directly to the model parameters and is the basis for the parameter estimator in Section 2.3, as well as for understanding and explaining the network topology and in-degree growth in terms of the fitted parameters.

Unfortunately, this theoretical model produces a *degree sequence*, not a graph. It predicts how many citations each node will accumulate, but it does not specify the edge list, the DAG structure, or the clustering patterns that emerge from the growth process.

**Algorithm 1** CS DAG Generation.

**Require:** $N$, $k$, $\{p_i, m_i, \rho_i, \sigma_i^2\}_{i=1}^{k}$
**Ensure:** Directed graph $G = (V, E)$ with community labels
Create $k$ seed nodes (one per community); $\mathrm{Urn}_i \leftarrow \emptyset$ for all $i$
**for** $v = k + 1$ **to** $N$ **do**
  $c(v) \leftarrow \mathrm{Categorical}(p_1, \dots, p_k)$; $c \leftarrow c(v)$ ▷ $c(\cdot)$ is the community-label map
  **if** $\sigma_c^2 > m_c$ **then** ▷ Overdispersed out-degree
    $\theta_c \leftarrow m_c^2/(\sigma_c^2 - m_c)$
    $\mu \sim \mathrm{Gamma}\big(\mathrm{shape} = \theta_c,\ \mathrm{scale} = (\sigma_c^2 - m_c)/m_c\big)$
    $d_{\mathrm{out}} \leftarrow \min(\mathrm{Poisson}(\mu), v{-}1)$
  **else** ▷ Fallback to Poisson distribution at low dispersion
    $d_{\mathrm{out}} \leftarrow \min(\mathrm{Poisson}(m_c), v{-}1)$
  **end if**
  $n_{\mathrm{acc}} \sim \mathrm{Binomial}(d_{\mathrm{out}}, 1 - \rho_c)$;
  $n_{\mathrm{pref}} \leftarrow d_{\mathrm{out}} - n_{\mathrm{acc}}$
  $T \leftarrow \emptyset$ (multiset)
  **for** $j = 1$ **to** $n_{\mathrm{acc}}$ **do** ▷ Accidental targets
    $T \leftarrow T \uplus \{\mathrm{Uniform}(1, \dots, v{-}1)\}$ ▷ We use $\uplus$ to denote multiset union
  **end for**
  **for** $j = 1$ **to** $n_{\mathrm{pref}}$ **do** ▷ Preferential targets
    **if** $|\mathrm{Urn}_c| > 0$ **then**
      $u \leftarrow \mathrm{Urn}_c\big[\mathrm{Uniform}(1, \dots, |\mathrm{Urn}_c|)\big]$
      $T \leftarrow T \uplus \{u\}$
    **else** ▷ Fallback to uniform sampling if no vertex in community $c$ has a positive in-degree
      $u \sim \mathrm{Uniform}[\text{community } c \setminus \{v\}]$
      $T \leftarrow T \uplus \{u\}$
    **end if**
  **end for**
  **for** each $u \in T$ (with multiplicity) **do**
    $E \leftarrow E \uplus \{(v, u)\}$;
    $\mathrm{Urn}_{c(u)} \leftarrow \mathrm{Urn}_{c(u)} \uplus \{u\}$
  **end for**
**end for**
**Output:** $E$ as a simple graph (prune duplicate edges) or as a weighted graph with weights corresponding to edge counts

CS bridges this gap: it implements the growth rules as an actual graph generator, adding the carefully engineered modifications described below to handle practical requirements that the analytical model abstracts away.

### 2.2. *From theory to the generator*

Each of the following adjustments is motivated by a specific gap between the theoretical model and real-world graph generation requirements. All modifications preserve the invariant that the expected out-degree of a vertex in community $i$ is $m_i$ ($\mathbb{E}[d_{\mathrm{out}}] = m_i$, exactly for all but the first few nodes of each community, where the clamp of Eq. (3) applies; see Section 2.2.1), ensuring that the theoretical predictions remain valid in expectation.

Algorithm 1 presents the complete DAG generation procedure. The algorithm produces a multi-graph, but duplicate edges can be pruned to produce a simple graph. Back-edge injection (Section 2.2.5) is applied as a post-processing step on the output of Algorithm 1 and is not shown to avoid clutter.

#### 2.2.1. *Stochastic out-degree with overdispersion*

The theoretical model assumes every node in community $i$ creates exactly $m_i$ edges. Real citation networks show substantial within-community variance in reference list lengths: some papers cite 3 references, others cite 80, even within the same field. We introduce a fourth per-community parameter $\sigma_i^2$ controlling out-degree variance.

When $\sigma_i^2 > m_i$ (the overdispersed case, typical in practice), we draw the out-degree from a Gamma-Poisson compound:

$$\begin{aligned} \theta_i &= m_i^2 / \left(\sigma_i^2 - m_i\right), \\ \mu &\sim \mathrm{Gamma}\left(\mathrm{shape} = \theta_i,\ \mathrm{scale} = \left(\sigma_i^2 - m_i\right)/m_i\right), \\ d_{\mathrm{out}} &= \min\big(\mathrm{Poisson}(\mu),\ v - 1\big). \end{aligned} \quad (3)$$

This is a negative binomial distribution with $\mathbb{E}[d_{\text{out}}] \approx m_i$ and $\text{Var}[d_{\text{out}}] = \sigma_i^2$. We sample it as a Gamma-Poisson compound rather than with a standard negative-binomial routine because the compound admits a real-valued shape $\theta_i$, whereas standard samplers require an integer stopping parameter. The $\min(\cdot, v-1)$ term caps the out-degree at the number of available predecessors – the $v-1$ nodes already present when node $v$ is added – which prevents anomalously dense connections during early growth (which is why the mean is only approximate).

When $\sigma_i^2 \leq m_i$ (the equi- or underdispersed case), we fall back to $d_{\text{out}} \sim \text{Poisson}(m_i)$, for which $\mathbb{E}[d_{\text{out}}] = \text{Var}[d_{\text{out}}] = m_i$. Appendix A.2 analyses how pruning duplicate edges affects the recovered parameters; the resulting underestimation of $m_i$ and $\sigma_i^2$ is measurable but negligible in practice.

### 2.2.2. Stochastic accidental/preferential split

The theoretical model deterministically allocates $(1-\rho_i)m_i$ accidental and $\rho_i m_i$ preferential edges. Since our $d_{\text{out}}$ is now stochastic, the split must also be stochastic. Given the realised out-degree $d_{\text{out}}$, we draw $n_{\text{acc}} \sim \text{Binomial}(d_{\text{out}}, 1-\rho_i)$ and set $n_{\text{pref}} = d_{\text{out}} - n_{\text{acc}}$. This preserves $\mathbb{E}[n_{\text{acc}}/d_{\text{out}}] = 1-\rho_i$ in expectation.

### 2.2.3. Urn-based preferential sampling

Naïve preferential sampling – computing all in-degrees, normalising, and drawing from the resulting distribution – costs $O(N_i)$ per edge, giving $O(NE)$ total generation time, where $N_i$ is the number of vertices in community $i$. We instead maintain a Pólya urn per community: whenever node $u$ receives an incoming edge, an additional copy of $u$ is appended to its community's urn. Drawing a preferential target reduces to sampling a uniformly random element from the urn, which is $O(1)$. The urn's composition is proportional to in-degree by construction. Total cost of the DAG generation phase is $O(N+E)$.

### 2.2.4. Cold-start handling

Early in the generation, two degeneracies arise from the graph still being small, both confined to the first few nodes of each community.

First, a community's urn is empty until one of its nodes receives an incoming edge, and preferential sampling from an empty urn is undefined. When this occurs, we fall back to uniform sampling over the community's existing members (excluding the new node, to prevent self-loops), which treats all members as equally attractive before any citations exist. This is a reasonable initialisation that converges to the true preferential distribution as the urn fills, and it is triggered only rarely – at most once per community. Second, a node may sample a target out-degree exceeding the number of nodes available to cite. The clamp $\min(\cdot, v-1)$ in Eq. (3) caps the out-degree at the $v-1$ existing predecessors, preventing anomalously dense connections while the graph is small. The first $k$ nodes (one per community) are created as seed nodes without outgoing edges, guaranteeing that no community is ever empty when it is first sampled.

### 2.2.5. Back-edge injection

The growth model produces a strict DAG: every edge points from a newer node to an older one. Real citation networks have a small fraction $r$ of back-edges: papers citing future work added during revision, patents referencing later filings, simultaneous publications with cross-references.

After DAG generation, we inject back-edges to match the input graph's back-edge ratio $r$. The number of injected edges is $n_{\text{back}} = \lfloor r \cdot |E|/(1-r) \rfloor$, ensuring that the resulting fraction of back-edges equals $r$.

When real timestamps are available (as is the case for the DBLP and OGBN-ArXiv datasets described in Section 4.1), the back-edge set is identified directly from the temporal ordering. Otherwise, we estimate $r$ using a greedy heuristic: score each node by out-degree minus in-degree, sort in descending order, and declare any edge violating this ordering as a back-edge. This runs in $O(N \log N + E)$ time.

CS's back-edge placement exploits the temporal ordering and community structure of the growth process:

- Each candidate back-edge $(u \to v)$ requires $u < v$ in CS's node-ID ordering (the older node cites the newer).
- Source-target pairs prefer small rank gaps (sampled via a geometric distribution with success probability $1 - \exp(-0.1) \approx 0.095$) and intra-community pairs with probability $\beta$, the back-edge locality, estimated from the input graph (from the temporal ordering when timestamps are available, and via the Feedback Arc Set (FAS) heuristic [41] otherwise; see Appendix A.1 and Section 4.6).
- Candidates are filtered against existing edges and previously accepted back-edges.
- Collisions are rejected and resampled using an $O(1)$ average-case hash-set lookup, eliminating the need for sorting or binary searches.

The back-edge injection phase operates in expected $O(N+E)$ time. This efficiency is achieved by using a hash-based set for duplicate detection (which takes $O(E)$ to build and provides $O(1)$ lookups) and an $O(N)$ single-pass community index for intra-community pair sampling. Combined with the $O(N+E)$ DAG generation phase, the total expected time complexity of CS generation is $O(N+E)$ with $O(N \log N)$ parameter estimation and an optional heuristic temporal ordering.

## 2.3. Parameter estimation

All $4k+1$ parameters are estimated in closed form from the input graph's adjacency matrix and community labels (Algorithm 2). No iterative optimisation, gradient descent, or cross-validation is required.

**Algorithm 2** CS Parameter Estimation.

**Require:** Directed graph $G = (V, E)$, community-label map $c : V \to \{1, \ldots, k\}$

**Ensure:** Per-community parameters $\{p_i, m_i, \rho_i, \sigma_i^2\}_{i=1}^{k}$ (the two global back-edge parameters $r$ and $\beta$ are estimated separately; Section 2.2.5)

1: **for** $i = 1$ **to** $k$ **do**
2: $V_i \leftarrow \{v : c(v) = i\}$; $N_i \leftarrow |V_i|$ ▷ Community $i$ and its size
3: $n_i \leftarrow N_i - 1$ ▷ No. of edge-emitting nodes: the seed emits none
4: $\hat{p}_i \leftarrow N_i / N$
5: $\Psi_i^{\text{out}} \leftarrow \sum_{v \in V_i} d_{\text{out}}(v)$
6: $\Psi_i^{\text{in}} \leftarrow \sum_{v \in V_i} d_{\text{in}}(v)$
7: $\hat{m}_i \leftarrow \Psi_i^{\text{out}} / n_i$ ▷ Expected out-degree per emitting node
8: $\hat{\sigma}_i^2 \leftarrow \frac{1}{n_i}\left(\sum_{v \in V_i} (d_{\text{out}}(v))^2 - n_i \hat{m}_i^2\right)$ ▷ Out-degree variance
9: $\gamma_i \leftarrow \text{Gini}\big(\{d_{\text{in}}(v) : v \in V_i\}\big)$ ▷ In-degree Gini
10: $\hat{\rho}_i \leftarrow \dfrac{\Psi_i^{\text{in}} (2\gamma_i + N_i - 2\gamma_i N_i)}{\Psi_i^{\text{out}} (\gamma_i + 1 - \gamma_i N_i)}$
11: **end for**

The four parameters per community have concrete interpretations:

- $p_i$: community growth probability – the field's share of new publications. Estimator: $N_i/N$ (direct count).
- $m_i$: expected out-degree – the typical reference list length in field $i$. Estimator: total out-degrees in community $i$ divided by $n_i = N_i - 1$, the number of nodes that emit edges. The input graph has no seed node; the correction is applied because the generator creates one edge-free seed per community, so only $N_i - 1$ of the $N_i$ generated nodes draw out-edges. The same $n_i$ is used in the variance estimator, so the two are consistent.
- $\sigma_i^2$: out-degree variance – heterogeneity of referencing habits within the field. Estimator: variance of out-degrees over the $n_i$ emitting nodes, using the same $n_i$ as $\hat{m}_i$.
- $\rho_i$: preferentiality – the fraction of citations that are merit-driven (proportional to existing in-degree) rather than accidental. Estimator: derived from the Gini coefficient of in-degrees via [31, Equation (19)]; see Step 10 of Algorithm 2. Intuitively, $\rho_i$ and the in-degree Gini index measure citation inequality from two sides: higher preferentiality concentrates citations on a few hubs, raising in-degree inequality and hence the Gini coefficient. The equation inverts this monotone relationship to recover $\rho_i$ from the observed Gini.

Let us state the parameter count explicitly. Per community, CS fits $p_i$, $m_i$, $\rho_i$ and $\sigma_i^2$; the growth probabilities carry one linear constraint, $\sum_i p_i = 1$, leaving $4k - 1$ free community parameters. Two further quantities are global and are also estimated from the input graph: the back-edge ratio $r$ and the intra-community back-edge locality $\beta$ (Section 2.2.5). The total is therefore $4k + 1$ for CS and $4k - 1$ for CS (DAG), which injects no back-edges. Both are $O(k)$; we report the exact counts rather than the leading term because the comparison in Section 5.6 uses them directly. The geometric rank-gap rate of Section 2.2.5 and the clamping tolerance $\epsilon$ are fixed constants, identical across all datasets, and thus omitted from the parameter count, and the graph size $N$ and community count $k$ are taken as given for every generator in the comparison.

The estimation procedure runs in $O(N \log N)$ time, dominated by the sorting of in-degrees for the Gini computation. The estimated $\hat{\rho}_i$ is clamped to $[\epsilon, 1 - \epsilon]$ with $\epsilon = 10^{-3}$ to ensure valid probabilities; clamping indicates the community may violate model assumptions. In practice, clamping is negligible on the smaller, well-formed networks (0–3% of communities on Cora, CiteSeer, Cora-Full, OGBN-ArXiv, and Cit-Patents) but non-trivial on the two datasets where those assumptions are most strained: PubMed (1 of 3 communities) and DBLP (≈24%), the latter being fragmented into tens of thousands of communities, many too small for reliable estimation.

When no input graph is available, all $4k + 1$ parameters can be specified directly. Every parameter has a concrete bibliometric interpretation, making it possible to construct synthetic networks from domain knowledge alone, for example, to generate a hypothetical citation network for a new field with specified growth rate, typical reference list length, and citation inequality. Appendix A.3 shows the one-at-a-time sweeps of the parameters and their corresponding network characteristics.

The consistency of this estimation, and the sensitivity of the generated structure to the intra-community back-edge locality $\beta$, are examined in Appendix A.1. In brief, $m_i$ and $\sigma_i^2$ are recovered consistently, $\rho_i$ carries a modest upward bias, and the reported metrics are largely insensitive to the locality parameter within its estimated range.

### *2.4. Theory validation*

Fig. 2 confirms that the CS's generated in-degree distributions match the theoretical predictions of Eq. (1) across three parametrisations spanning the full range of preferentiality.

At $\rho = 0.2$ (low preferentiality, where most citations are accidental), the generated complementary CDF closely tracks the Pareto type II prediction across the entire degree range. At $\rho = 0.5$ (moderate preferentiality), theory and generation remain aligned over three orders of magnitude in degree. At $\rho = 0.9$ (high preferentiality), the bulk of the distribution matches well, with visible deviation only in the extreme tail. This is expected: at high $\rho$, a small number of hub nodes accumulate disproportionate in-degree, and finite-size effects ($N = 30{,}000$) cause the empirical tail to fluctuate around the asymptotic prediction. These deviations diminish with increasing $N$, which is consistent with the theoretical convergence to Pareto type II.

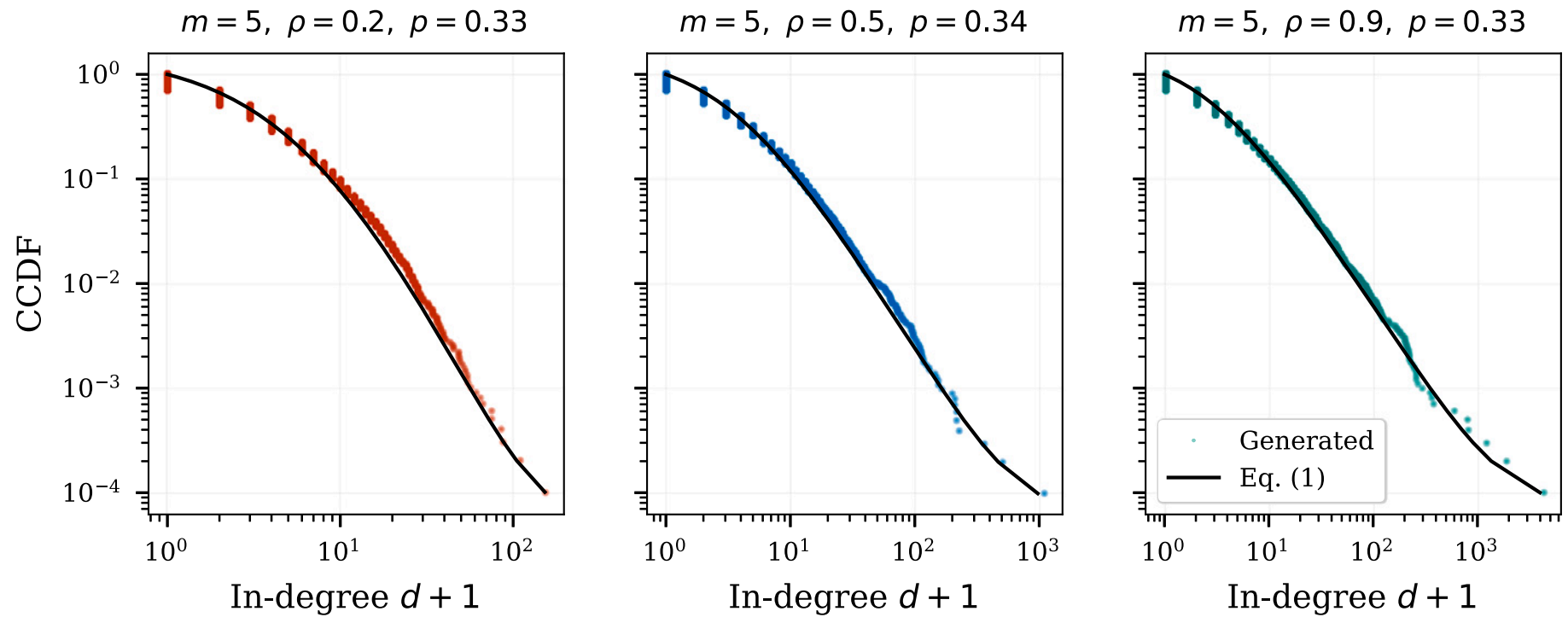

**Fig. 2.** CS-generated vs theoretical in-degree distributions (Eq. 1), $N = 30{,}000$, $k = 3$, and $\rho = 0.2$ (low preferentiality, left), $\rho = 0.5$ (moderate, centre) or $\rho = 0.9$ (high, right). The theoretical CCDF is computed from the same parameters. Deviations in the tails reflect finite-size effects that diminish with increasing $N$. All three communities are equal-sized; the 0.33/0.34/0.33 distinction is for display purposes only.

This validation is important because the theoretical CCDF is *not fitted* to the generated data: both the curve and the generator use the same parameter values. This agreement confirms that the engineered modifications described in Section 2.2 (stochastic out-degree, urn-based sampling, cold-start handling) do not distort the theoretical degree structure in expectation. The evaluation on 26 metrics in Section 5 will show that this fidelity extends well beyond the in-degrees, to structural properties the model was never designed to reproduce. That evaluation concerns a static snapshot; the separate question of extrapolation over time is addressed, for the in-degree distribution only, in Appendix A.4.

The agreement in Fig. 2 is not coincidental. The only network parameters on which the expected in-degrees depend are the accidental rate $\langle a \rangle$ and the effective preferentiality $\nu_i$ (Eq. 1), which are functions of $\{p_i, m_i, \rho_i\}$ alone. CS introduces three elements absent from the analytical model, and each preserves these quantities in expectation for large enough $v$:

1. the stochastic out-degree draw satisfies $\mathbb{E}[d_{\text{out}}] = m_i$ for all but a handful of the first nodes due to clamping (Section 2.2.1),
2. the random accidental/preferential split $n_{\text{acc}} \sim \text{Binomial}(d_{\text{out}}, 1-\rho_i)$ gives $\mathbb{E}[n_{\text{acc}}] = (1-\rho_i)m_i$,
3. the urn sampler draws preferential targets with probability proportional to in-degree by construction, reproducing the attachment kernel of Brzozowski et al. [31] exactly.

Consequently, CS simulates a process governed by the same master equation as the analytical model, and Eq. (1) is its solution for the expected in-degree of a node at local time $t$. We stress that Eq. (1) describes in-degree *growth* and is not itself a stationary quantity: it depends explicitly on $t$. The stationary object is its $t \to \infty$ limit, the Pareto type II law of Section 2.1. The clamping of $d_{\text{out}}$ (Eq. 3) occurs in a negligible number of turns in practice. Deduplication to a simple graph is the only other step that may perturb the realised degrees, and Appendix A.2 shows this effect to be similarly minor.

## 3. Cycle breaking in other models

In this section we define and motivate the process we call "cycle breaking", i.e., an adaptation of existing graph generators to the structure of citation-like networks that are sparse in cycles. Previously known static generators do not operate under the assumption of this sparsity. Instead, they copy and recreate estimated parameters and patterns without global constraints. As such, we first consider the process and explain the choices made.

### 3.1. The process

The particular process we propose for adapting a graph generator's output into a citation-like network structure is simple; we describe it in the points below.

- Read the input graph's timestamps to determine the chronological order of node appearance; if the timestamps are unavailable, use a greedy heuristic that sorts nodes by $d_{\text{out}} - d_{\text{in}}$ to determine the approximate order. Using the order found, compute the fraction of back-edges (i.e., going from older nodes to newer nodes) as $r$.
- Using a similar greedy heuristic process, assign a topological ordering to the generator's output nodes. Then, reverse all edges to point from newer nodes to older nodes, creating a DAG.
- Select a fraction $r$ of edges from the newly created DAG at random and reverse them to create back-edges.

### 3.2. Considered alternatives

The above process, while intuitive and simple, should be meticulously scrutinised to make sure that our proposition for cycle breaking is preferably unbiased and offers a fair ground for comparison of different generators:

*Decycling first.* It would be plausible to skip the decycling step and try to obtain the ratio of $r$ back-edges directly from the generator's output. While it is possible, it necessarily poses the question "which edges to reverse" since the topological order, at this point, cannot be easily deduced; we believe any strategy selected would be biased toward the specific generator, and we hope to achieve a model-agnostic procedure that can be compared between various generators. As such, we believe that beginning with the decycling step is a reliable solution for creating a methodologically correct starting point to achieve a truthful comparison.

*Reversing vs addition.* An alternative to the proposed approach would be adding back-edges instead of reversing existing edges. One could argue that by reversing the direction of existing edges, we interrupt the flow patterns induced by each generator. To this, we say that we have already interrupted this flow by enforcing the DAG structure, which is a necessary sacrifice for further consistency. We also argue that if, at this point, we added edges instead of reversing them, we introduce unnecessary bias caused by the variability of $r$ in different input graphs. By reversing edges, we exactly preserve the skeletal structure, sacrificing only the flow patterns of the graph.

*Generator input.* There is another possibility: first remove back-edges from the input graph, fit the generator to produce the output, and then reintroduce a fraction of $r$ edges as back-edges. The difference between this approach and the above one is that the generated graphs are fitted to the DAG subgraph of the input instead of the noisy full version. We have considered this approach, but we believe it introduces a similar bias to that described before: a fraction of $r$ edges in the output graph are not only reversed, but also placed against the generator's output. This would, in our opinion, be a *de facto* new hybrid model instead of a neutral modification of the baseline. Also, neither of the considered reference algorithms was designed with the imitation of a DAG network in mind, so we see no methodological benefit in fitting the model to the input graph stripped of back-edges.

With the exhaustive explanation of key choices above, we hope to offer a fair ground for the comparison of the effects of cycle breaking on reference algorithms.

## 4. Experimental setup

This section describes the evaluation infrastructure shared by all analyses in the paper: the datasets, the generators under comparison, the structural metrics, and the statistical methodology.

### 4.1. Datasets

We evaluate all methods on seven directed networks with ground-truth community labels (Table 2). These span three orders of magnitude in node counts (from Cora with 2.7k nodes to DBLP with 3.6M M) and community counts ranging from 3 (PubMed) to 63k (DBLP), providing diversity in both scale and structural complexity. To the best of our knowledge, these seven networks represent the most comprehensive publicly available collection, because the available real-world networks satisfying all required conditions are simply very rare.

Here are some brief notes on individual datasets. **Cit-Patents** [42] is the only patent network in our benchmark suite; it is a patent citation graph with USPTO patent classifiers as communities. **DBLP** [43] is the largest by node/edge count and community count; it is a paper-to-paper citation graph derived from the AMiner v14 database, with field-of-study metadata as community labels (the field of science with the highest weight was selected as the community label of each node). **OGBN-ArXiv** [44] is a medium-scale paper citation graph (169 K nodes) with ArXiv category labels, widely used as a machine learning benchmark. **Cora**, **CiteSeer**, and **PubMed** are smaller paper-to-paper standard benchmarks with field-of-study labels [30]. **Cora-Full** [45] is a lesser-known, more comprehensive variant of the first of these (Cora). All datasets were downloaded and processed using the UnPACD repository [46]. The Cit-Patents and DBLP datasets contained substantial fractions of nodes with missing community labels. For our analysis, they were removed from the graphs.

### 4.2. Generators under comparison

We compare 12 methods organised into three groups: baseline generators, community-aware generators, and CS variants. Every algorithm from the literature was considered in two versions: standard and adjusted with cycle breaking; see Table 3 for a summary.

*Five classical generators (standard variants).* We include Erdős-Rényi (ER) [47], the Configuration Model [47] (abbreviated as Config Model in tables and figures, and when naming its variants), the Stochastic Block Model (SBM) [48], the Degree-Corrected SBM (DC-SBM) [11], and the LFR benchmark [13]. Each is fitted to an input graph's relevant statistics: edge density for ER, degree sequence for the Configuration Model, block connection matrix and community assignments for SBM and DC-SBM, and power-law exponents with the mixing parameter for LFR. These generators produce undirected or densely cyclic directed graphs; none natively outputs a DAG or near-DAG.

**Table 2**
Datasets used in the evaluation. Datasets marked with an asterisk (*) were subsampled for the purpose of computing computationally expensive metrics (see Section 4.5). The value $N$ corresponds to the number of nodes after pruning vertices without community labels, where required.

| Dataset | $N$ | $\hat{d}_{\text{out}}$ | $k$ | Timestamps |
|---|---|---|---|---|
| Cit-Patents* [42] | 2,755,865 | 5.07 | 418 | Discarded |
| CiteSeer [30] | 3312 | 1.39 | 6 | Absent |
| Cora [30] | 2708 | 2.00 | 7 | Absent |
| Cora-Full [45] | 19,793 | 3.30 | 70 | Absent |
| DBLP* [43] | 3,586,177 | 6.24 | 62,904 | Used |
| OGBN-ArXiv* [44] | 169,343 | 6.89 | 40 | Used |
| PubMed [30] | 19,717 | 2.25 | 3 | Absent |

**Table 3**
Generator comparison across capability dimensions. "Near-DAG" indicates native support for a low back-edge-ratio directed acyclic topology; generators lacking it require the cycle breaking transformation of Section 3 (the "-nD" variants). Parameter counts are for $k$ communities on an $N$-node graph; for networks such as DBLP the $k^2$ term heavily dominates. Generation complexity refers to the graph-generation phase.

| Model | Directed | Native near-DAG | Community-aware | No. of parameters | Generation complexity |
|---|---|---|---|---|---|
| ER | Yes | No | No | $O(1)$ | $O(N+E)$ |
| Config Model | Yes | No | No | $O(N)$ | $O(N+E)$ |
| LFR | Yes | No | Yes | $O(1)$ | $O(N+E)$ |
| SBM | Yes | No | Yes | $O(k^2)$ | $O(N+E)$ |
| DC-SBM | Yes | No | Yes | $O(N+k^2)$ | $O(N+E)$ |
| **CS** | Yes | **Yes** | Yes | $O(k)$ | $O(N+E)$ |

*Five classical generators (near-DAG variants).* Each standard generator is also evaluated after applying the cycle breaking described in Section 3. The transformation imposes a heuristic topological ordering on the generated graph, reverses the direction of edges to follow this ordering (producing a DAG), then reintroduces a fraction $r$ of edges as uniformly random back-edges, where $r$ is the ratio of back-edges. The resulting variants are denoted by a "-nD" suffix (e.g., DC-SBM-nD). This doubling of the baseline count is deliberate: it allows us to evaluate cycle breaking's impact as a methodological question in its own right (Section 3), independent of the CS comparison.

*CS and CS (DAG).* The iterative generator CS was described in Section 2. It produces a directed near-DAG with planted community labels using $4k+1$ analytically estimated parameters ($k$ = the number of communities): growth probability $p_i$, expected out-degree $m_i$, preferentiality $\rho_i$, and out-degree variance $\sigma_i^2$ for each community $i$, plus the two global back-edge parameters $r$ and $\beta$, less the constraint $\sum_i p_i = 1$ (Section 2.3). The CS (DAG) variant omits the back-edge injection step, producing a strict DAG.

### 4.3. Structural metrics

We evaluate structural fidelity by comparing each synthetic graph against the corresponding real graph across a battery of metrics. Three distance measures are used, chosen to match the nature of each metric. In all cases, lower values indicate better fidelity; zero indicates a perfect match:

- **Absolute Percentage Error (APE)** for scalar graph properties (e.g., average path length, global clustering coefficient, modularity): APE = $|x_{\text{synth}} - x_{\text{real}}|/|x_{\text{real}}|$.
- **Wasserstein 1-distance ($W$)** for distributional properties (e.g., in-degree distribution, betweenness distribution, community-size distribution): the minimum transport cost to transform one distribution into the other.
- **$L_1$ distance** for the triad census (16 triad types), which has no canonical real-line ordering suitable for the Wasserstein distance: $L_1 = \sum_\tau |f_{\text{synth}}(\tau) - f_{\text{real}}(\tau)|$, where $\tau$ indexes the 16 triad types and $f(\cdot)$ is the corresponding proportion.

The 26 metrics[1] are organised into six structural categories, each capturing a distinct aspect of graph structure. Each metric belongs to exactly one category:

1. **Global Topology** (3 metrics): APE of the effective diameter, APE of the average path length, $W$ of the reachability distribution. It captures large-scale connectivity and distance structure.
2. **Degree Structure** (4 metrics): $W$ of in-degree distribution, $W$ of out-degree distribution, APE of in-degree assortativity, APE of out-degree assortativity. Captures degree distributions and degree-degree correlations.

[1] We started with 27 candidate metrics, and checked possible redundancy with the pairwise Spearman rank correlation $r_s$: for each pair of metrics with $|r_s| \geq 0.85$ (computed on within-cell standardised values, pooled across all datasets and methods), we retained only one. This procedure removed only a single metric (APE of ground-truth assortativity, redundant with other community-level metrics), yielding 26 retained metrics.

3. **Mesoscopic Endogenous (Structure)** (6 metrics): APE of ground-truth modularity, APE of ground-truth conductance, APE of ground-truth inter-cluster edge density, APE of ground-truth intra-cluster edge density, $W$ of ground-truth in-degree participation, $W$ of ground-truth out-degree participation. This captures how well the synthetic graph reproduces the statistical properties of the community structure to which the graph is assigned by real labels; see Section 4.4 for discussion.
4. **Mesoscopic Exogenous (Structure)** (6 metrics): APE of Infomap codelength, APE of Leiden modularity (with standard modularity quality function), APE of Leiden modularity (with RBConfiguration quality function), $W$ of Infomap community-size distribution, $W$ of Leiden community-size distribution (modularity), $W$ of Leiden community-size distribution (RBConfiguration). This captures how community-detection algorithms *behave* on the synthetic graph compared to the real input one; see Section 4.4 for discussion.
5. **Local Structure** (4 metrics): APE of global clustering coefficient, APE of feed-forward loop count, $W$ of local clustering coefficient distribution, $L_1$ of triad census. It captures clustering patterns, motifs, and triadic structure.
6. **Flow & Connectivity** (3 metrics): $W$ of the betweenness centrality distribution, $W$ of the strongly connected component size distribution, $W$ of the longest-path-length distribution. It captures directed flow, component structure, and DAG depth. Back-edges were omitted when computing longest paths for tractability.

### *4.4. The endogenous/exogenous distinction*

The partition of mesoscopic metrics into endogenous and exogenous categories requires some more detailed justification. We use the word “endogenous” to refer to the ground-truth partition vector supplied with each graph, whilst “exogenous” indicates a partition found by an external classifier such as a community detection algorithm; similar usage can be found in Legramanti et al. [49].

*Endogenous mesoscopic metrics.* We compare the ground-truth community structure of the synthetic graph against that of the real graph, answering whether *the planted communities in the synthetic graph share the statistical properties of the real ones*. These metrics take both the graph and its community labels as input. While these metrics are widely accepted as measures of concordance between input and generated graphs, we argue they are inherently biased by the interaction between *externally imposed* topology (from ground-truth labels) and the graph’s *intrinsic* topology (natural density fluctuations). Consider two extremes. In some citation networks, these two align cleanly: intra-community connections dominate inter-community ones, making the graph a prime candidate for SBM fitting. In others, a community label is merely *indicative* of a node’s local connections rather than determinative – it may describe a paper’s citation tendencies without defining its full local structure. Such networks call for modelling tools better suited than a plain SBM. More fundamentally, evaluating a generator on these metrics rewards similarity in the *relationship* between topology and community labels across input and output, rather than similarity between the graphs themselves. We do not claim this is worthless – but for downstream tasks such as community detection, it can mislead us into “believing” the generator faithfully imitates the input’s structure when it has only matched a derived property.

*Exogenous mesoscopic metrics.* Run the standalone community-detection algorithms (Infomap [50], Leiden [51]) on both the synthetic and real graphs and compare the resulting partitions. They answer the question of whether *the community-detection algorithms behave the same way on synthetic graphs as they do on real graphs*. These metrics take only the graph as input; they are “blind” to the planted labels. We argue that this way of comparing graphs that are to serve as community detection benchmarks is better-fitted to the task. By comparing the performance of established solutions, we can check if the synthetic outputs present, for example, the same biases on both input and output graphs, without the need to consult external and potentially noisy “declared” labels.

A caveat applies to all exogenous metrics: they inherit the inductive biases of the detector used. We deliberately employ two algorithms from different paradigms – Infomap (map equation) and Leiden (modularity optimisation) – so that the exogenous comparison does not privilege a single algorithmic notion of “community”. A method that mispartitions the real and synthetic graphs in the same systematic way could mask a real difference, while one unusually sensitive to a particular motif could exaggerate it. As such, the results presented should be considered only relative to the chosen algorithms, and adding spectral or inference-based community detection algorithms is a natural future direction.

A broader caveat applies to the evaluation as a whole. Any finite battery of structural metrics is a human construction: it encodes design choices and cannot, by itself, prove that a synthetic graph is indistinguishable from a real one on every conceivable downstream task. We mitigate this in two ways. The metrics are chosen to span six complementary structural categories and are filtered for pairwise redundancy (footnote 1); and, more importantly, the exogenous metrics move beyond summary statistics to a behavioural test – whether community detection algorithms act on the synthetic graph as they do on the real one. We regard the latter as the closest available proxy, within a static-snapshot benchmark, to the task-based validation that a fully dynamic evaluation would demand. Crucially, no generator in our comparison is tuned toward these metrics: each is configured only from closed-form estimators on the input graph and is blind to the benchmark objective, so a high score reflects fidelity rather than optimisation against the measurement.

The endogenous/exogenous distinction is not specific to citation networks. It requires only two ingredients: a planted ground-truth labelling, and an algorithm whose behaviour can be compared between the real and synthetic graphs. Any generative benchmark for a label-based task therefore admits it – undirected or attributed community-detection benchmarks, and, more broadly, node-classification or link-prediction benchmarks, where one can likewise separate “does the synthetic graph reproduce the planted statistic” (endogenous) from “does a real algorithm behave on it as it does on the true graph” (exogenous). Two conditions govern how useful the split is elsewhere. First, the exogenous side presupposes a canonical reference algorithm: community detection offers de facto standards (Infomap, Leiden), whereas a domain lacking a settled method makes the exogenous comparison more detector-dependent, an amplified form of the caveat already noted above. Second, the distinction is informative only where planted labels and intrinsic topology can diverge; in benchmarks whose labels are, by construction, nearly identical to the structure, the two metric families largely coincide. Its value is greatest precisely where labels are indicative rather than determinative, e.g., in the regime in which citation networks sit.

### *4.5. Statistical pipeline*

Our evaluation employs a practical pipeline for comparing all algorithms whilst hoping to retain some statistical elegance. All randomness derives from a single master seed, making the benchmark fully reproducible; seeds do not overlap across methods, ensuring the ensembles are independent for the statistical tests.

*Stage 1: Friedman omnibus test.* For each dataset, the 12 methods are ranked from 1 (best) to 12 (worst) based on their mean distance across 50 independent samples for all retained metrics. The Friedman test checks whether the observed mean ranks are consistent with the null hypothesis that all methods perform identically. It reports the $\chi^2$ statistic and the corresponding $p$-value; rejection indicates that at least one method differs significantly.

*Stage 2: Mann-Whitney U tests.* Since the results in each block (dataset, metric) contain potentially dependent values (because, e.g., average path lengths are necessarily correlated with longest path lengths), we cannot run standard statistical procedures using all 182 (dataset, metric) blocks as independent samples. On the other hand, with only 7 datasets we do not expect the Nemenyi critical difference diagram or the Wilcoxon signed-rank test to indicate significant differences between any pair of algorithms. Instead, we compare the 50 results obtained by each pair of algorithms on each (dataset, metric) cell and report wins and ties as computed using the Mann-Whitney $U$ test with uncorrected $p < 0.05$. While these task-level counts are not substitutes for global significance claims, they provide a rigorous, descriptive breakdown of exactly where and how frequently each model outperforms the baselines. To remove the impact of scale, all values in a given (dataset, metric) cell are standardised before aggregation. This has no impact on method ranks.

To quantify uncertainty in the mean rank estimates, we additionally report 95% bootstrap confidence intervals for each method's mean rank. The intervals are obtained by block-resampling (dataset, metric) rows with replacement (10,000 draws) and recording the 2.5th and 97.5th percentiles of the resulting distribution of mean ranks. Because this procedure resamples the atomic evaluation unit – the (dataset, metric) block – it correctly propagates both dataset-level and metric-level variance without assuming independence across blocks.

*Large-graph handling.* For the three largest datasets (Cit-Patents, DBLP, OGBN-ArXiv), several metrics are computationally prohibitive on the full graph. We compute these on subgraphs of up to 50,000 nodes obtained via BFS from random seed nodes. Computationally heavy metrics (e.g., average path length, betweenness centrality) are estimated from 2000 sampled node pairs or paths. The same subsampling procedure is applied identically to both the real and synthetic graphs, ensuring that any approximation error cancels in the distance computation.

### *4.6. Evaluation design*

Table 4 summarises the experimental design.

The five classical generators are evaluated in two configurations: standard (as published) and near-DAG (Section 3). This doubling is not an artefact of the comparison but a deliberate design choice that allows us to address two questions simultaneously: (1) how do the generators compare to each other, and (2) how does cycle breaking affect each generator's fidelity? Together with CS and CS (DAG), this gives $M = 12$ methods.

All methods receive the same input: the real graph's adjacency matrix and community labels (where applicable). CS and the block models (SBM, DC-SBM) use community labels; the Configuration Model and ER use only degree or density statistics; LFR uses global structural parameters estimated from the real graph. This heterogeneity in input is inherent to the methods and is reflected in the parsimony analysis (Section 5.6). For two datasets, DBLP and OGBN-ArXiv, real timestamps are used when searching for the set of back-edges. The remaining datasets are first equipped with an approximate topological ordering based on the greedy FAS heuristic [41], sorting nodes in descending order by the difference between their out- and in-degrees. The Cit-Patents network has missing timestamps for nearly $1/3$ of the nodes, so the real timestamps were discarded in favour of the heuristic as well. Let us also note that there exists an asymmetry in back-edge placement between CS and cycle breaking in the reference methods that results from one process being model-specific and the other model-agnostic; to alleviate potential issues in comparison, we separate the analysis of the impact of adding back-edges to CS from adding them to the reference methods.

**Table 4**
Summary of the experimental design.

| Component | Count |
|---|---|
| Datasets | 7 |
| Methods (5 base × 2 variants + 2 CS) | 12 |
| Metrics | 26 |
| Metric categories | 6 |
| Samples per method per dataset | 50 |
| (dataset, metric) blocks | $7 \times 26 = 182$ |
| (dataset, metric) blocks (non-endogenous) | $7 \times 20 = 140$ |

## 5. Results

We organise the results of our benchmark into six subsections. We first present the global ranking of all 12 methods (Section 5.1), then dissect it by structural category (Section 5.2) and by the endogenous/exogenous distinction (Section 5.3) introduced in Section 4.4. Section 5.4 isolates the contribution of the back-edge mechanism through an ablation study. Section 5.5 examines per-dataset variation to identify where different methods succeed and fail. Finally, Section 5.6 relates performance to parameter complexity.

### 5.1. Global comparison

Figs. 3, 4, and Table 5 summarise the global ranking of all 12 methods. The Friedman omnibus test [52] rejects the null hypothesis that all methods perform identically ($\chi^2 = 22.6$, $p = 0.02$, 7 datasets).

For convenience, we group the methods into three descriptive tiers. High-parameter descriptive models and CS form the top tier, outperforming mid-tier structure-agnostic models, while models with constant parameter counts sit at the bottom.

Several observations deserve emphasis. First, the top tier contains both high-parameter descriptive models (DC-SBM-nD with $O(N+k^2)$ parameters) and low-parameter models (CS with $O(k)$), suggesting that parameter count alone does not determine structural fidelity on this task. Second, cycle breaking improves nearly every baseline's mean rank, with improvements ranging from minor (Configuration Model, LFR) to notable (DC-SBM, SBM, ER). Third, we observe that models with a constant number of parameters (ER, LFR) perform the worst. This indicates that the complex nature of citation networks cannot be properly encapsulated by a small set of values.

The Δ column in Table 5 records the rank shift when endogenous metrics are excluded (Section 5.3). A consistent pattern emerges: every method that encodes ground-truth community structure worsens, while every method without it improves or remains stable. This systematic pattern motivates the non-endogenous analysis in Section 5.3.

### 5.2. Category-level analysis

The global rank averages results over structurally diverse metrics and inevitably masks important category-level variation. Fig. 5 and Table 6 decompose performance into six structural categories. Category scores are defined as $s = (M + 1 - \bar{r})/M$, where $\bar{r}$ is the method's mean rank within the category and $M = 12$ is the number of compared methods. The central finding is that no single method dominates all categories. The category leaders are:

- **Global Topology:** Growth-based DAG construction and degree-preserving rotation both produce realistic distance/diameter structure.
- **Degree Structure:** DC-SBM and Configuration Model both preserve the input's degree sequence by construction and are expected to dominate this category; CS's growth process approximates it from community-dependent parameters.
- **Meso. Endogenous:** The top models directly encode the ground-truth block structure with $O(N+k^2)$ parameters. SBM-nD and SBM also score well, reflecting their explicit community parametrisation.
- **Meso. Exogenous:** DC-SBM-nD drops to 0.520, noticeably below the methods it dominates on endogenous metrics. Here SBM variants, Configuration Model, and CS take the lead. That DC-SBM-nD reverses here is the empirical basis for the endogenous vs exogenous distinction: memorising the ground-truth community structure does not guarantee that algorithms will perceive the synthetic graph similarly to the real one.

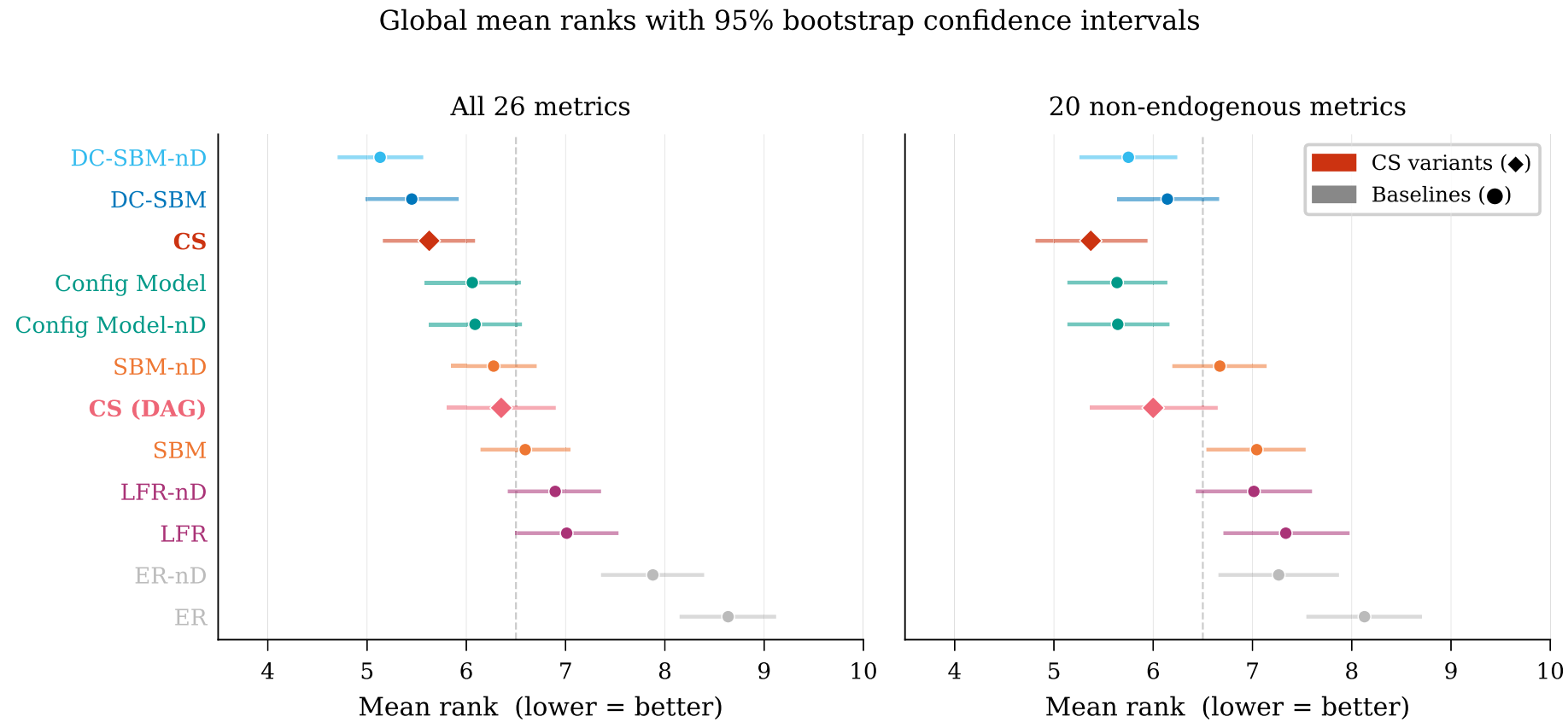

**Fig. 3.** Global mean ranks with 95% bootstrap confidence intervals, obtained by block-resampling of (dataset × metric) rows with replacement ($B = 10{,}000$ draws). Point estimates are shown as diamonds (◆) for CS variants and circles (•) for baselines; horizontal bars span the 2.5th-97.5th percentiles of the bootstrap distribution. The dashed vertical line marks the median rank (6.5) for $M = 12$ methods. Left panel: all 26 metrics (182 blocks); right panel: 20 non-endogenous metrics (140 blocks).

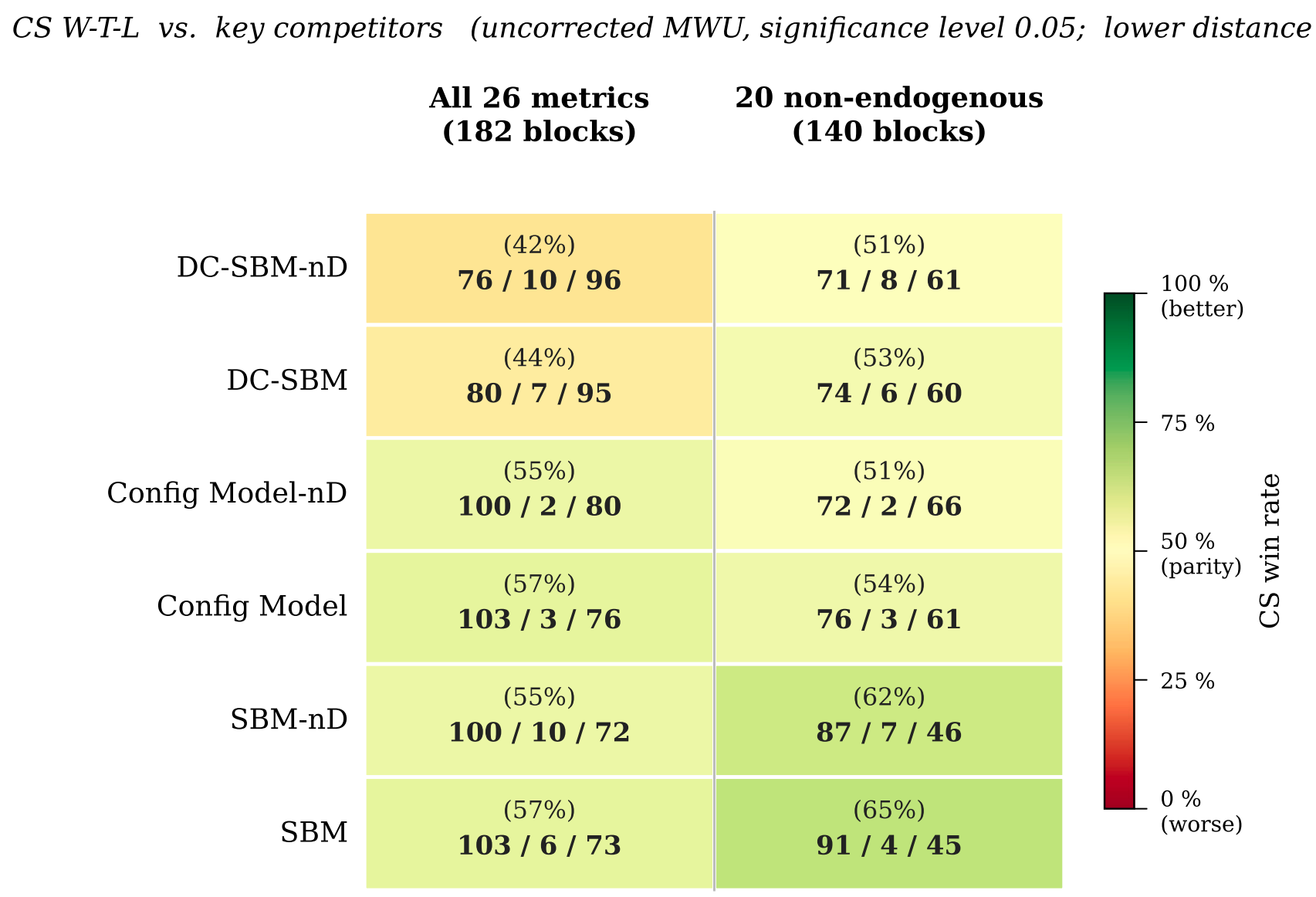

**Fig. 4.** Pairwise comparison: number of CS wins/ties/losses against its six nearest competitors, computed via uncorrected Mann-Whitney $U$ test (significance level 0.05) based on the 50 runs per (dataset, metric) block. Cell colour encodes CS's win rate (green > 50% = CS advantage; red < 50% = competitor advantage). Left column: all 26 metrics (182 blocks); right column: 20 non-endogenous metrics (140 blocks).

**Table 5**
Mean ranks across 26 retained metrics (182 blocks) and the 20 non-endogenous subset (140 blocks), with 95% bootstrap confidence intervals obtained by block-resampling (dataset × metric) rows with replacement ($B$ = 10,000 draws). Δ is the rank change when endogenous metrics are excluded (negative = improvement). Sorted by Rank (26). See Fig. 4 for pairwise comparisons.

| Method | All 26 metrics | | 20 non-endogenous | | |
|---|---|---|---|---|---|
| | Rank | 95 % CI | Rank | 95 % CI | Δ |
| DC-SBM-nD | 5.13 | [4.70, 5.57] | 5.75 | [5.26, 6.24] | +0.62 |
| DC-SBM | 5.45 | [4.98, 5.92] | 6.14 | [5.64, 6.66] | +0.69 |
| **CS** | 5.63 | [5.16, 6.09] | 5.37 | [4.81, 5.94] | −0.25 |
| Config Model | 6.06 | [5.58, 6.55] | 5.64 | [5.14, 6.14] | −0.42 |
| Config Model-nD | 6.09 | [5.62, 6.56] | 5.64 | [5.14, 6.16] | −0.45 |
| SBM-nD | 6.27 | [5.85, 6.71] | 6.67 | [6.19, 7.14] | +0.40 |
| **CS (DAG)** | 6.35 | [5.80, 6.90] | 6.00 | [5.36, 6.65] | −0.35 |
| SBM | 6.59 | [6.14, 7.05] | 7.04 | [6.54, 7.54] | +0.45 |
| LFR-nD | 6.90 | [6.42, 7.36] | 7.01 | [6.43, 7.60] | +0.12 |
| LFR | 7.01 | [6.49, 7.53] | 7.34 | [6.71, 7.98] | +0.32 |
| ER-nD | 7.88 | [7.36, 8.40] | 7.26 | [6.66, 7.87] | −0.61 |
| ER | 8.64 | [8.15, 9.12] | 8.13 | [7.54, 8.71] | −0.51 |

- **Local Structure:** Degree-preservation coupled with granular inter- and intra-cluster density reconstruction is likely to force similar local patterns as in the input graph, driving the notable lead over the competition. We note the good performance of CS despite no parameters directly encoding local structure of nodes.
- **Flow & Connectivity:** The Configuration Model's exact degree preservation combined with topological rotation produces highly realistic flow structure. DC-SBM-nD scores only 0.464: static block models have no native mechanism for generating meaningful flow.

The category landscape reveals a fundamental trade-off. High-parameter models with explicit community encoding (DC-SBM, SBM) dominate endogenous mesoscopic metrics but underperform on flow and connectivity. Structure-agnostic models (Configuration Model) excel at degree and flow fidelity but do not have the community awareness required to replicate the input's heterogeneity. Growth-based models (CS, CS (DAG)) occupy a firm middle ground, with competitive performance across all categories and excelling in replicating the challenges that input graphs pose for community detection algorithms (Meso. Exogenous). CS is also notable for its *balance*: its minimum category score is 0.544 (Meso. Endogenous), making it, among the eight methods shown in Table 6, the only one whose worst category exceeds 0.47. This balance means CS is a reasonable default when the downstream task's structural requirements are not known in advance.

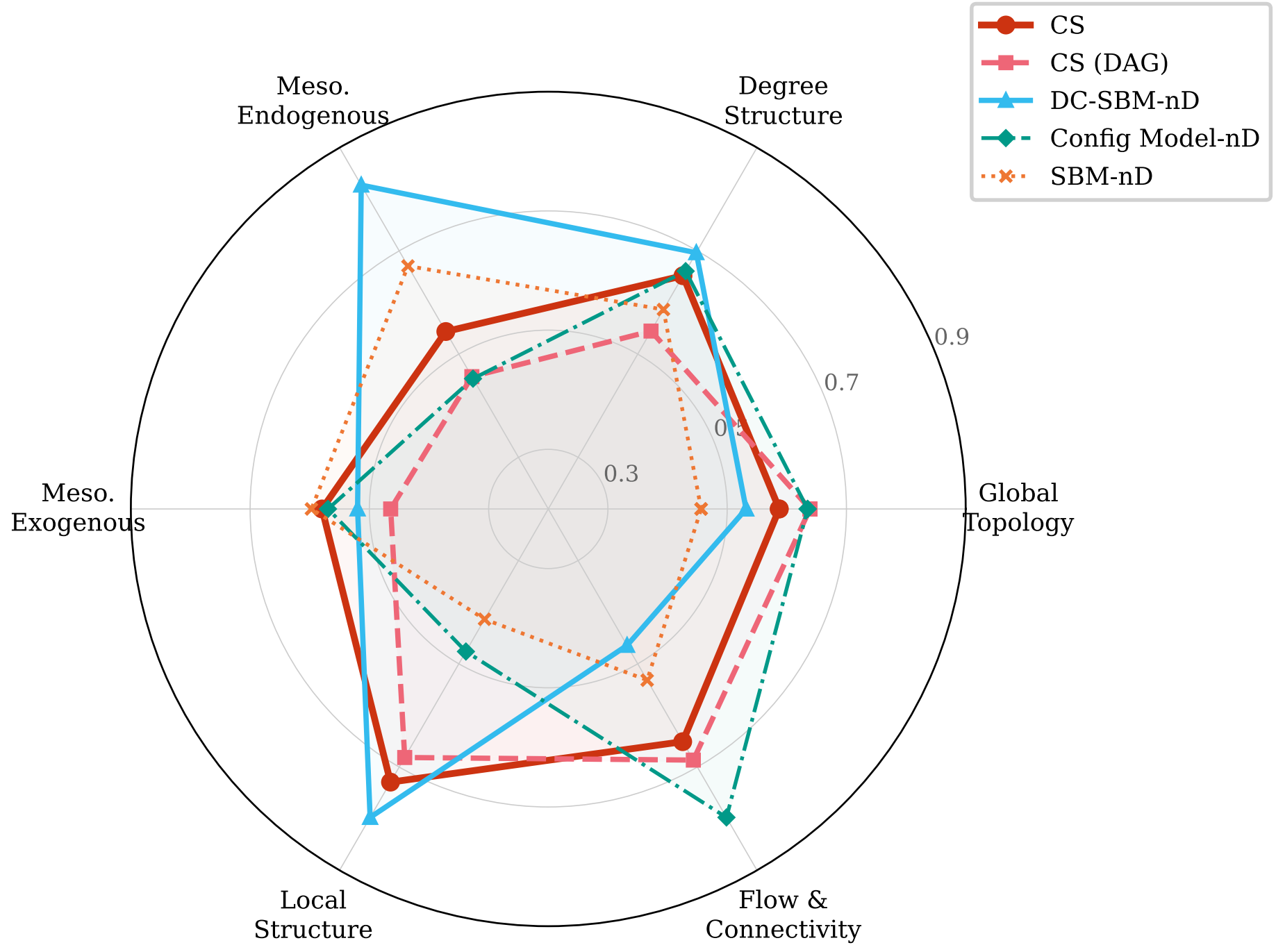

**Fig. 5.** Category-level performance scores for selected methods. Score is $= (M + 1 - \bar{r})/M$, where $\bar{r}$ is the category mean rank and $M = 12$ is the number of methods. Higher is better; 0.5 indicates median performance.

**Table 6**
Category-level performance scores (higher = better): $(M + 1 - \bar{r})/M$, where $M = 12$ is the number of compared methods.

| Method | Global Topology | Degree Structure | Meso. Endogenous | Meso. Exogenous | Local Structure | Flow & Connectivity |
|---|---|---|---|---|---|---|
| **CS** | 0.587 | 0.652 | 0.544 | 0.579 | 0.729 | 0.651 |
| **CS (DAG)** | 0.639 | 0.545 | 0.456 | 0.464 | 0.681 | 0.686 |
| DC-SBM-nD | 0.532 | 0.696 | 0.827 | 0.520 | 0.798 | 0.464 |
| DC-SBM | 0.508 | 0.539 | 0.821 | 0.571 | 0.759 | 0.429 |
| Config Model-nD | 0.635 | 0.661 | 0.452 | 0.569 | 0.476 | 0.798 |
| Config Model | 0.607 | 0.694 | 0.460 | 0.593 | 0.461 | 0.758 |
| SBM-nD | 0.456 | 0.586 | 0.671 | 0.597 | 0.414 | 0.532 |
| SBM | 0.389 | 0.521 | 0.659 | 0.625 | 0.429 | 0.405 |

### *5.3. Non-endogenous analysis*

In this section, we turn our attention back to the Δ column in Table 5 and the 20-metric ranks already reported there. When the 6 endogenous metrics are excluded, the ranking shifts substantially for several methods:

- DC-SBM-nD worsens from 5.13 to 5.75 ($\Delta = +0.62$);
- DC-SBM worsens from 5.45 to 6.14 ($\Delta = +0.69$);
- Config Model-nD improves from 6.09 to 5.64 ($\Delta = -0.45$);
- CS improves from 5.63 to 5.37 ($\Delta = -0.25$).

On the 20 non-endogenous metrics, CS (5.37) and the two Configuration Model variants (both 5.64) occupy the top three positions, while DC-SBM-nD drops to fourth place (5.75). The mean-rank gap between CS and DC-SBM-nD reverses in CS's favour, from 0.50 against CS among all 26 metrics to 0.38 for CS on the non-endogenous subset. Notably, on the 42 evaluation blocks removed, the Win–Tie–Loss record of DC-SBM-nD against CS was 35–2–5; the largest advantage of DC-SBM-nD therefore lay in precisely this group, and over the remaining blocks the record of DC-SBM-nD against CS is 61–8–71.

The systematic nature of the shift is more informative than any individual gap. Every method with explicit community parametrisation (DC-SBM, DC-SBM-nD, SBM, SBM-nD) worsens when endogenous metrics are removed. Every method without it (Configuration Model, ER, CS) improves or remains stable. The exception here is LFR, which also degrades despite encoding only general parameters. These observations are consistent with the hypothesis that endogenous metrics reward memorisation of planted statistics, while exogenous metrics reward functional realism: the ability to produce a graph on which community-detection algorithms behave as they do on the real network. Both perspectives are valid evaluation criteria; we argue that for the specific use case of community-detection benchmarking, exogenous metrics are the primary concern.

### 5.4. Ablation: cycle breaking

Let us consider how cycle breaking impacts the reference methods' performance in each category, and how removing back-edges impacts CS.

Fig. 6 exposes three patterns. First, Mesoscopic Exogenous fidelity is the sole universal casualty: every baseline degrades there, with four of the five markedly (29–36%) and ER marginally (48%). Imposing and partially reversing a global topological order dissolves the short-range feedback loops and reciprocal motifs that Infomap and Leiden exploit to delineate communities, so a detector perceives a different mesoscopic structure on the transformed graph than on the real one – even as large-scale statistics improve. Second, the size of the benefit tracks how far the base generator is from a DAG. The generators furthest from it gain most on Degree Structure (DC-SBM 89%, LFR 82%), whereas the Configuration Model, which already reproduces the degree sequence, has only a downside there (36%) and is the sole baseline that is net-harmed overall. Third, the trade-off aligns with a clean structural axis: cycle breaking consistently improves long-range and degree properties, which a coherent topological ordering makes directly visible to the generator, while it erodes the mesoscopic-community properties that depend on local cyclic coverage. Local Structure sits between the two, reflecting that triadic patterns are driven partly by long-range order and partly by community structure.

To isolate the contribution of community-aware back-edge injection, we compare CS against CS (DAG); see Fig. 7 and Table 7. The overall rank Δ is 0.72, which indicates a practical difference and shows clear category-specific patterns.

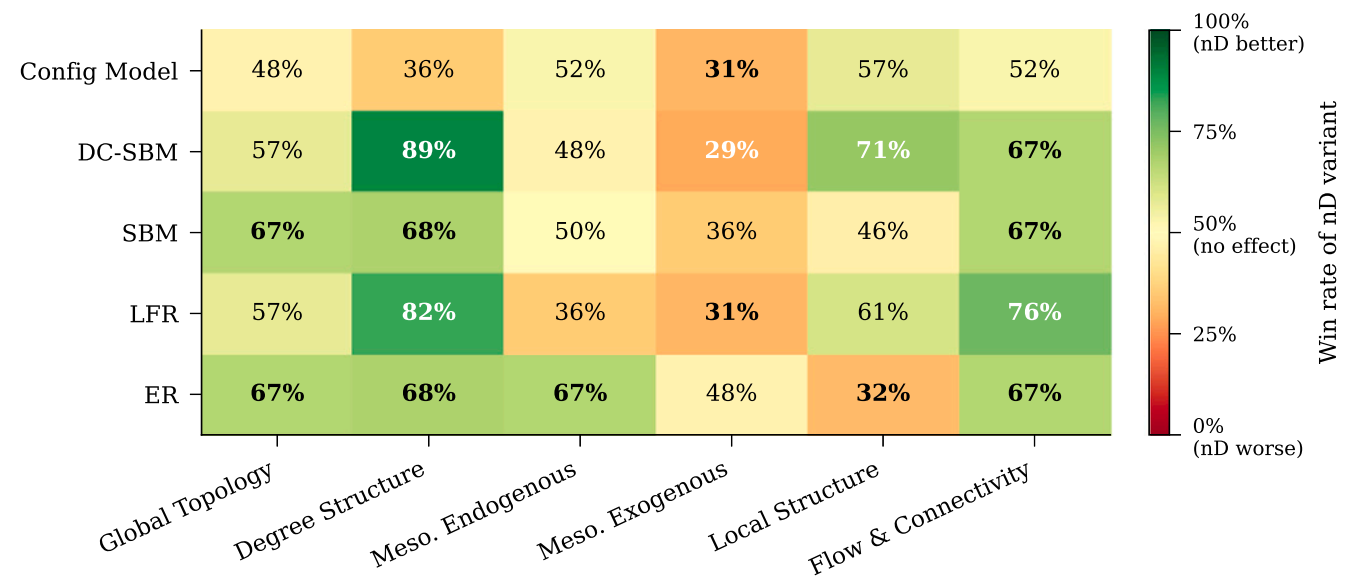

**Fig. 6.** Heatmap presenting the impact of cycle breaking per category. Each cell represents a percentage of instances in which the near-DAG version of the method achieved better results than the standard method on the Corresponding metrics.

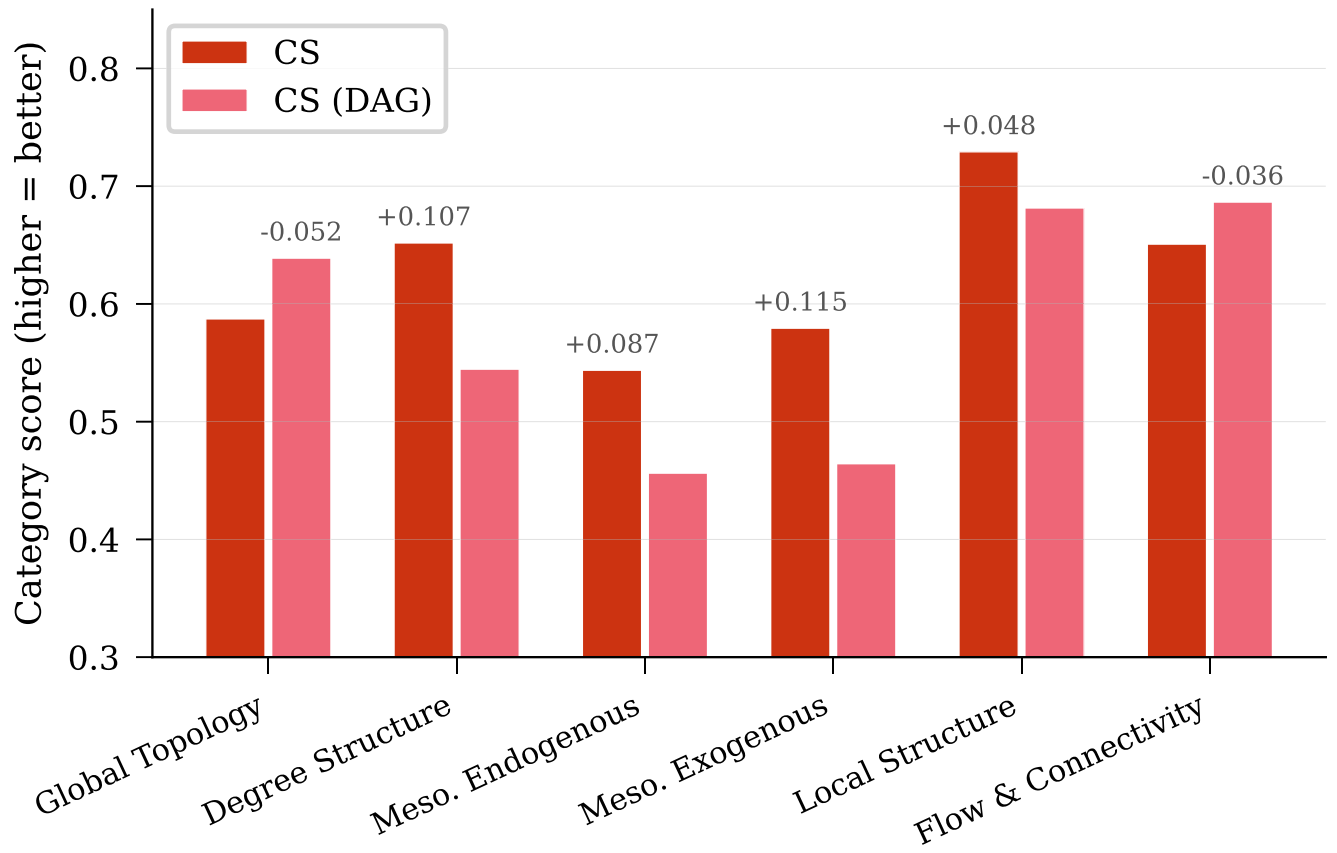

**Fig. 7.** Ablation: effect of back-edge injection on category scores. Positive differences (annotated) indicate categories where CS outperforms the pure DAG variant.

**Table 7**
Ablation: CS vs. CS (DAG). "DAG wins" counts the (dataset × metric) cells in which CS (DAG) achieves the lower distance; CS wins the complementary cells. Back-edges improve degree structure and mesoscopic fidelity at the cost of global topology and, to a lesser extent, flow and connectivity. The score columns are as defined in Section 5.2.

| Category | DAG wins | Score (CS) | Score (DAG) |
|---|---|---|---|
| Global Topology | 15/21 | 0.587 | **0.639** |
| Degree Structure | 9/28 | **0.652** | 0.545 |
| Meso. Endogenous | 18/42 | **0.544** | 0.456 |
| Meso. Exogenous | 12/42 | **0.579** | 0.464 |
| Local Structure | 8/28 | **0.729** | 0.681 |
| Flow & Connectivity | 10/21 | 0.651 | **0.686** |

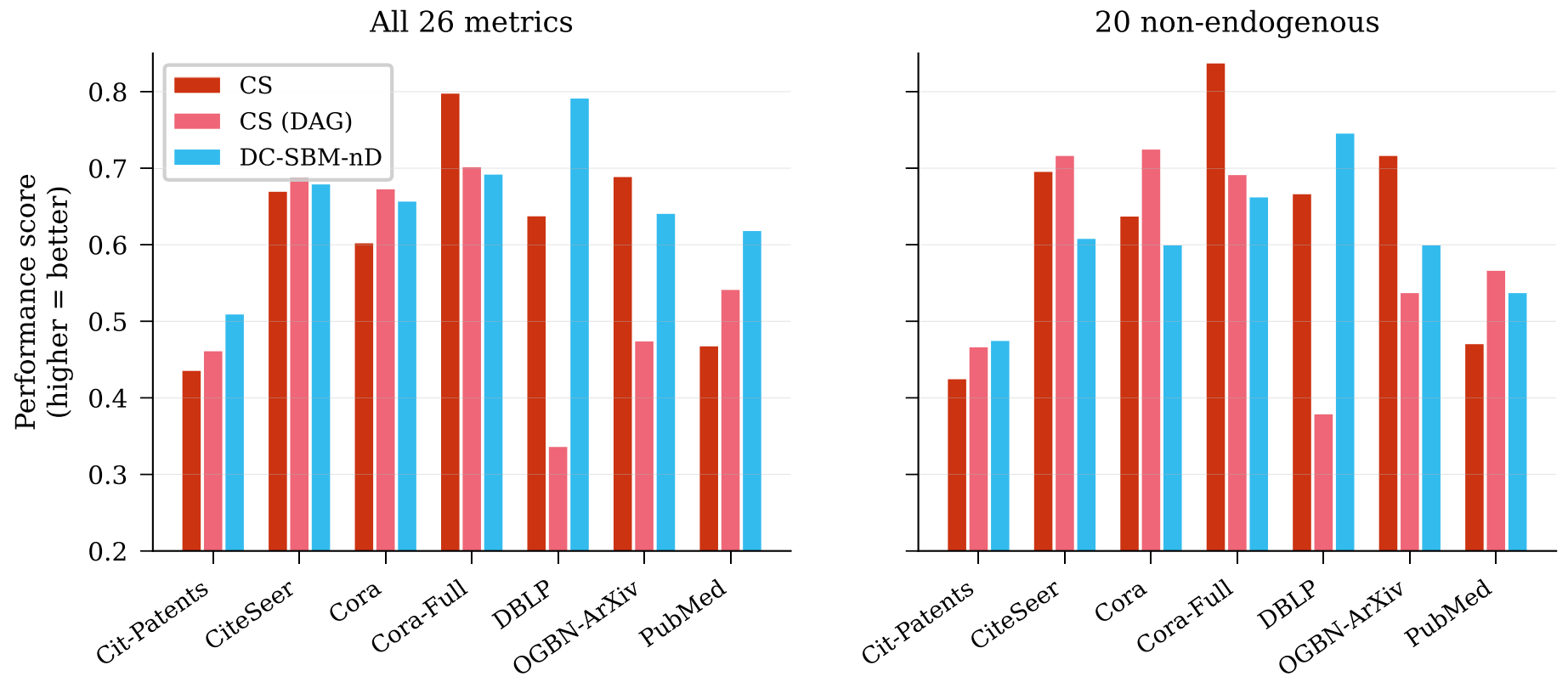

**Fig. 8.** Per-dataset performance scores for CS, CS (DAG), and DC-SBM-nD. Left: all 26 metrics; right: 20 non-endogenous metrics. Score = $(M + 1 - \bar{r})/M$ averaged within each dataset.

Back-edges help in four categories. The largest effect is on Meso. Exogenous, where CS wins in 30 of 42 dataset×metric cells: the short-range cycles introduced by back-edges make community-detection algorithms perceive the graph more realistically. Degree Structure (19/28), Meso. Endogenous (24/42), and Local Structure (20/28) also improve, because back-edges add the small feedback loops that are frequent in citation networks (small cycles, revision-induced references).

Back-edges hurt in two categories: Global Topology degrades noticeably (CS wins only 6/21), as back-edges shorten effective diameters and average path lengths, which the estimated parameters aimed to reproduce. Flow & Connectivity moves the same way in score, though the cell counts are near parity (CS wins 11/21), so the effect there is weak.

This trade-off has a practical implication: the choice between the variants should be task-driven. For community-detection benchmarking, CS with back-edges is preferred. For tasks sensitive to flow structure (influence propagation, temporal ordering), CS (DAG) is likely the better choice. Notably, this trade-off mirrors the pattern observed in the cycle breaking of baselines (Section 3): any transformation that introduces or removes cycles shifts the balance between mesoscopic/local fidelity and flow fidelity.

### 5.5. *Per-dataset variation*

Now we shall focus on the comparison of our algorithm to the main competitor, i.e., DC-SBM-nD. Fig. 8 shows the performance of the three methods.

The most striking pattern is that *relative performance depends on dataset characteristics*. CS shows an advantage on Cora-Full and OGBN-ArXiv – medium-scale networks with moderate community counts (70 and 40) where the Price-Pareto assumptions hold well. DC-SBM-nD exhibits better performance on DBLP and PubMed. DBLP is the largest network in our evaluation, where the uniform inter-community assumption becomes most problematic; PubMed has only 3 communities, limiting the Price-Pareto model's degrees of freedom. On Cit-Patents – the second largest network – DC-SBM-nD leads, but the difference in score is small, demonstrating that scale alone does not determine CS's performance; rather, the interaction between scale and community heterogeneity is the key factor. Notably, the theoretical work that introduces the model on which CS is built also proposes a variant taking into account the ageing of the nodes, although we omit it due to the need for numerical estimation of parameters. As we observe, though, CS remains highly competitive even on large networks where the ageing effect is typically observed. We attribute this effect to the fact that if an ageing effect is present, the parameter estimators may adjust for the ageing effect by outputting $\rho_i$ lower than the preferentiality really driving the network growth. Moreover, attention decay substantially reshapes in-degree distributions only over long horizons, and this effect may not be observable at full scale on the considered networks. We consider Citation Seeder's fidelity in forecasting in-degree growth in Appendix A.4. On the small benchmarks (Cora, CiteSeer), DC-SBM-nD and CS are comparable, with neither showing a consistent advantage.

The non-endogenous panel (right) shifts the picture. DC-SBM-nD beats both variants of CS only on one dataset (DBLP). We believe this effect may stem from the uniform accidental citation rate assumption being the most violated on the DBLP network, which contains a large number of very small communities. These results also confirm that DC-SBM-nD's overall edge in the full evaluation is partly attributable to its endogenous advantage, which is expected given its direct encoding of ground-truth community statistics.

### 5.6. *The parsimony argument*

The preceding sections establish *what* the methods achieve; this section asks what they achieve *relative to their complexity*. Fig. 9 plots the mean rank against the geometric mean of parameter counts across the seven datasets.

The left panel (26 metrics) reveals that the top-cluster methods occupy vastly different positions on the parameter axis. DC-SBM-nD (5.13) and DC-SBM (5.45) require $O(N + k^2)$, Configuration Model variants require $O(N)$, CS (5.63) requires $4k + 1$: 29 on Cora, 1673 on Cit-Patents, 251,617 on DBLP. The ratio between the parameter counts of DC-SBM-nD and CS exceeds $1700 : 1$ on Cit-Patents and $15{,}700 : 1$ on DBLP.

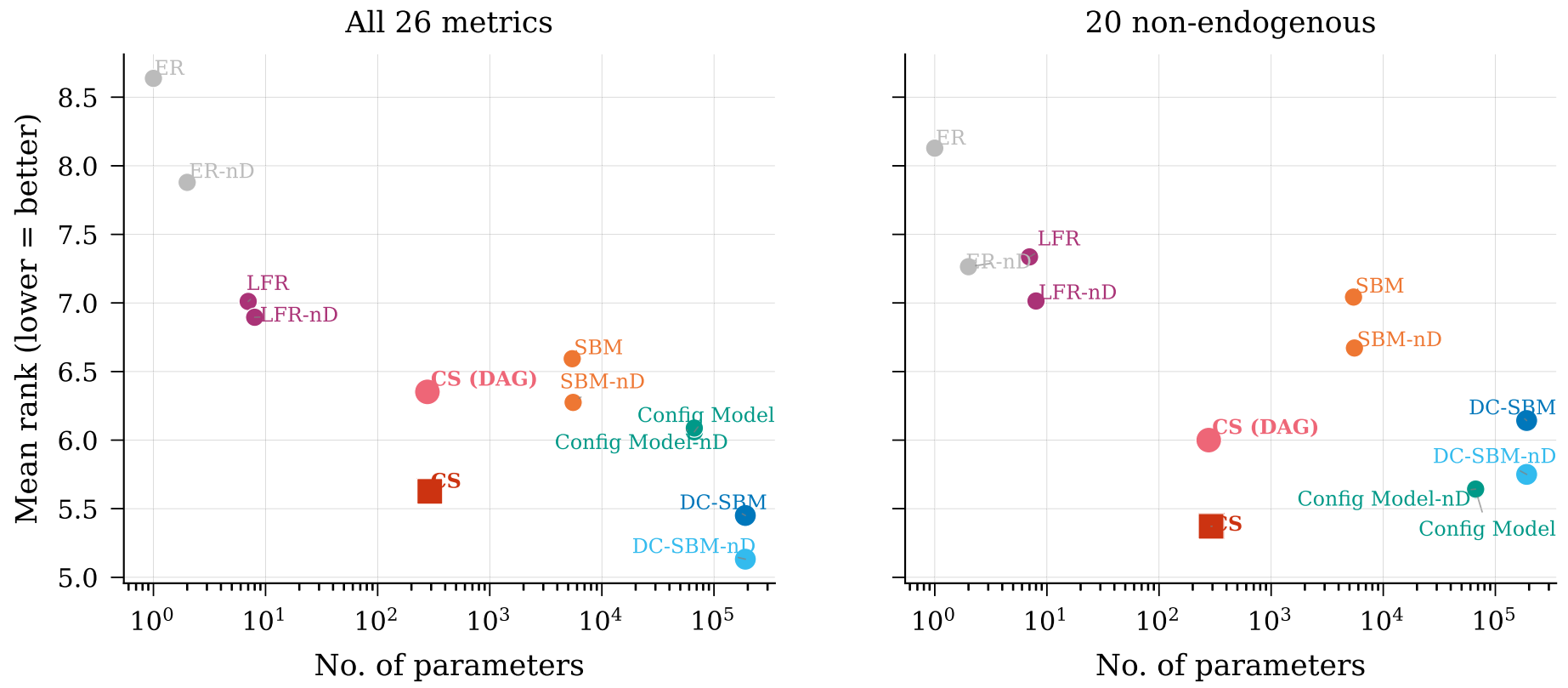

**Fig. 9.** Performance-parsimony trade-off. Horizontal axis: geometric mean of parameter counts across the 7 datasets (log scale). Vertical axis: mean rank (lower is better). Left: all 26 metrics; right: 20 non-endogenous metrics. The "-nD" variants carry one parameter more than their base generators, the back-edge ratio $r$; CS's count includes both $r$ and the back-edge locality $\beta$.

The right panel (20 non-endogenous) sharpens the picture. CS (5.37) moves into the Pareto-optimal position: no method achieves both a better rank and fewer parameters. Config Model-nD (5.64) achieves a similar rank with $O(N)$ parameters, but it produces no community structure – an essential requirement for community-detection benchmarking. Among community-aware methods, CS uniquely combines competitive fidelity with minimal parametrisation.

This parsimony gap is not merely a storage difference. It reflects a qualitative distinction between *descriptive* models that parametrise a specific graph and *mechanistic* models that parametrise a growth process. DC-SBM stores each node's expected degree and each block's connection probability – that is, the realised quantities the endogenous metrics measure – and its parametrisation grows with the graph. CS stores $p_i$, $m_i$, $\rho_i$, $\sigma_i^2$: community-level tendencies of how citations are produced, whose number is set by the number of fields rather than by the size of the network. We do not claim that a smaller parameter count is advantageous in itself – ER and LFR are the most parsimonious models here and rank last, and SBM attains the highest Meso. Exogenous score of all. The distinction we draw is that a descriptive parametrisation reproduces the graph it was fitted to, whereas a mechanistic one can be run forward to a larger horizon (Appendix A.4), specified from domain knowledge with no input graph at all (Section 2.3), and varied one interpretable knob at a time (Appendix A.3). The former can reproduce one graph faithfully; the latter can generate a family of structurally similar graphs from explicit assumptions about how citations occur.

## 6. Conclusion

The synthetic generation of directed, nearly acyclic graphs with planted communities has long been an underserved area in information sciences, severely limiting our ability to benchmark algorithms on citation-like topologies. In this work, we addressed this gap through three primary contributions: the introduction of the CS generator, the evaluation of cycle breaking as a transformation applied to static reference generators, and a separate validation in terms of endogenous and exogenous mesoscopic structures with detailed results. We believe our contributions shed new light on the task of generating nearly acyclic directed graphs: while DC-SBM remains most competitive in the task of network replication, it is not dominant in specific categories and underdelivers when it comes to replicating community detection algorithm performance. Moreover, CS reproduces that quality while matching exogenous mesoscopic structure more closely, and does so with far fewer parameters. Our analysis also shows that cycle breaking as a process generally helps all algorithms fit the input citation networks better, but it comes at the price of worse community recreation, which is an important caveat when evaluating community detection. Notably, these observations could not have been made were it not for the dichotomous methodology we employed for validating mesoscopic fits.

Two claims should be kept distinct. As a static generator fitted to a snapshot, CS attains top-tier fidelity across the 26 structural metrics using $O(k)$ parameters against DC-SBM's $O(N + k^2)$, and is Pareto-optimal on the non-endogenous subset (Section 5.6). Parsimony here is not a concession but a feature: an interpretable growth process, each of whose parameters carries a bibliometric meaning, matches the generative fidelity of models that instead parametrise the graph itself. As a growth model, however, CS's validated predictive scope is narrower. Appendix A.4 establishes that it extrapolates the future in-degree distribution more faithfully than both DC-SBM variants do on a single timestamped network, with the advantage widening as more history is observed; we make no claim about the future evolution of the remaining structural properties. Establishing that would require a dynamic benchmark over several timestamped near-DAGs – only two of our seven networks carry usable timestamps – so we leave the temporal evaluation of Citation Seeder to future work.

We should also be clear about CS's theoretical and methodological boundaries. First, the current Price-Pareto model lacks a mechanism for attention decay. Consequently, CS's in-degree fidelity degrades on massive networks with deep temporal histories. Integrating attention decay is theoretically possible, but doing so requires numerical parameter estimation, thereby sacrificing the analytical tractability that is CS's primary advantage. Second, the model assumes that accidental citations are distributed uniformly across all communities. In reality, interdisciplinary citation patterns are highly heterogeneous (e.g., machine learning papers cite statistics much more frequently than they cite history). Allowing for non-uniform inter-community citation rates would accurately capture this, but it would require an $O(k^2)$ parameter matrix, converging toward the parameter bloat of standard block models. Relatedly, CS plants a hard, birth-assigned partition rather than modelling emergent community formation. This is a deliberate design choice for a controllable generator with known ground truth; the datasets themselves typically supply only single-field labels, and cross-community citation already enters through the accidental channel. However, real citation networks are dynamic, and community labels could change with network growth.

Methodologically, our evaluation is constrained by the scarcity of suitable datasets. With only seven networks representing the comprehensive collection of directed near-DAGs with ground-truth communities, our statistical power is inherently limited; mean rank differences of 0.3 to 0.5, while directionally informative, cannot reach formal significance. We have considered two additional data sources that were ultimately omitted: OpenAlex (no consensus in the research community about which subgraph to take as the representative) and the APS citation network (limited availability and usage restrictions). Moreover, the methods receive different input information (labels for CS and the block models, degree sequence for the Configuration Model, density for ER, global parameters for LFR), so the rankings reflect both modelling capacity and information granted. The exogenous metrics neutralise the block models' label advantage, and the parsimony analysis reads quality relative to the parameters consumed.

These limitations naturally motivate our future work, and the two most immediate model extensions have contrasting feasibility profiles. Node ageing can reuse the attention-decay variant already derived for the underlying model: generation remains $O(N + E)$, since decay is a time-dependent reweighting of the community urns, but the closed-form Gini-based estimator of $\rho_i$ no longer applies, so the decay parameters must be fitted numerically – the challenge is preserving interpretability once estimation becomes iterative. Heterogeneous inter-community citation has the opposite profile: the scalar accidental rate $\langle a \rangle$ would be replaced by a $k \times k$ affinity matrix estimated in closed form from inter-community edge counts, cheap to fit but costing $O(k^2)$ additional parameters, so a low-rank or block-sparse approximation targeting the dominant cross-field flows is the natural compromise. In both cases the tension is the same – richer dynamics against the analytical parsimony that is CS's central advantage. Beyond these, we believe that Citation Seeder's fidelity to the structural properties evaluated here lets it serve as a controlled experimental ground, supporting studies that a limited selection of real-world datasets cannot.

## CRediT authorship contribution statement

**Łukasz Brzozowski:** Writing – review & editing, Writing – original draft, Visualisation, Software, Methodology, Investigation, Formal analysis, Data curation, Conceptualization. **Marek Gagolewski:** Writing – review & editing, Writing – original draft, Supervision, Methodology, Investigation, Conceptualization. **Grzegorz Siudem:** Writing – review & editing, Writing – original draft, Methodology, Investigation, Formal analysis, Conceptualization.

## Software and data availability

The CS generator is released as open-source software on https://github.com/lukaszbrzozowski/CitationSeeder.
The reproduction pipeline is available at https://github.com/lukaszbrzozowski/CDSI_DAG.
The UnPACD software used to process all datasets is available at https://github.com/lukaszbrzozowski/unpacd.

## Declaration of Generative AI and AI-assisted technologies in the writing process

During the preparation of this work, the first author used Claude (Anthropic) for the code generation of analysis scripts. The authors reviewed and edited all output and take full responsibility for the content of this publication.

## Declaration of competing interest

The authors declare that they have no known competing financial interests or personal relationships that could have appeared to influence the work reported in this paper.

## Appendix A. Supplementary results

This appendix provides further context for understanding Citation Seeder's fidelity and limitations. The following sections consider, in turn, how structural metrics respond to the back-edge locality parameter, whether the closed-form estimators recover the parameters that generated a graph, how parameter changes affect observable network characteristics, and how Citation Seeder compares with DC-SBM in forecasting in-degree growth.

### *A.1. Robustness analysis*

Fig. A1 probes how sensitive the generated structure is to $\beta$, the fraction of back-edges placed inside communities and the one back-edge parameter that is estimated indirectly. We fix a reference setting: $k = 10$ equal-sized communities of 1000 nodes each, $m_i = 8$, $\rho_i = 0.6$, $\sigma_i^2 = 16$, and back-edge ratio $r = 0.10$ – and we sweep $\beta$ over $\{0.5, 0.6, \dots, 1.0\}$ together with the data-estimated value of $\approx 0.72$, generating 10 near-DAGs per value and reporting ensemble means. The left panel normalises each metric to its value at 0.72; the right panel re-estimates $\beta$ from a generated graph, once from the true node ordering and once from the timestamp-free FAS heuristic, and plots the estimate against the injected value.

Across this range, modularity, the intra-community edge fraction and the in-degree Gini vary by at most a few percent, so the reported metrics are largely insensitive to the parameter. Reciprocity is the sole exception, changing by tens of percent – expected, since $\beta$ directly governs short intra-community cycles. In the right panel, the true ordering recovers the locality close to identity line, whereas the FAS heuristic used for timestamp-free graphs systematically underestimates it; we therefore estimate $\beta$ from timestamps when present.

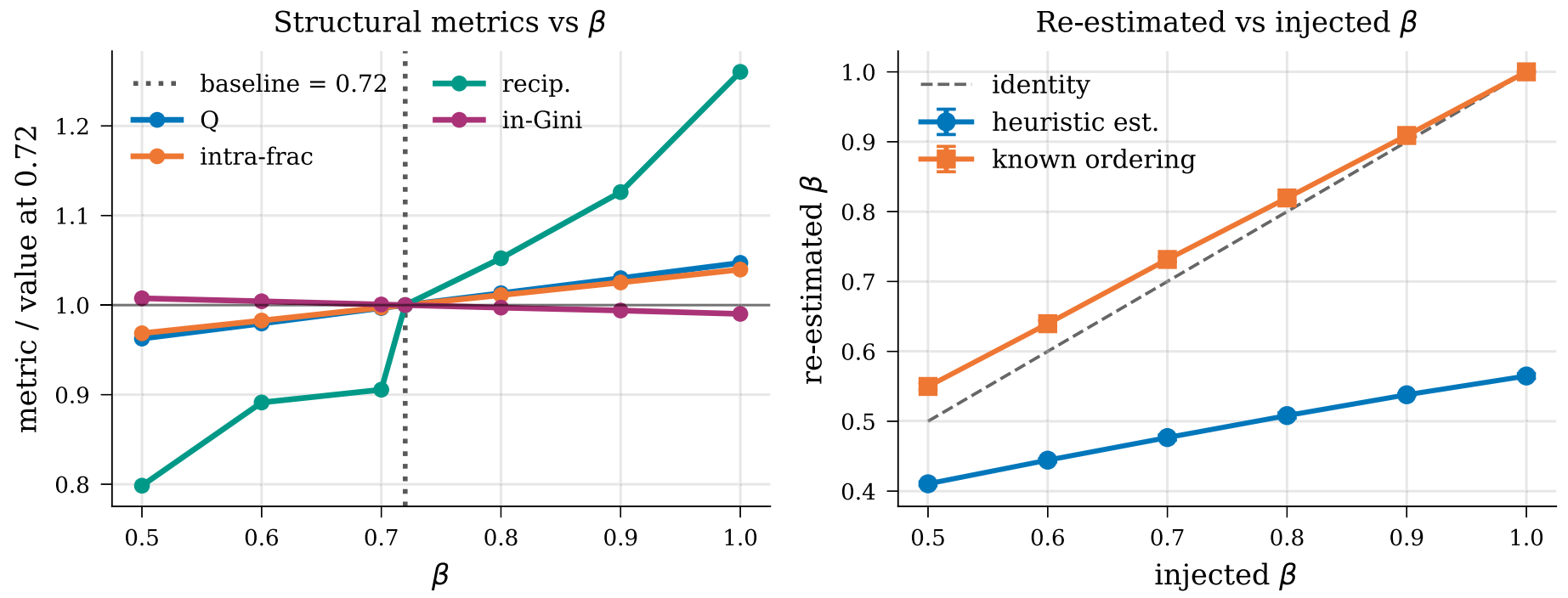

**Fig. A1.** Sensitivity to the intra-community back-edge locality ($\beta$) on a synthetic near-DAG. Left: four structural metrics as a function of $\beta$, each normalised to its value at the data-estimated operating point (dotted line, ≈0.72). Right: $\beta$ re-estimated from a generated graph against the injected value, using either the true node ordering or the timestamp-free FAS heuristic. Standard deviation-based error ranges on the right plot are too small to be visible.

### A.2. Estimator recovery

Figs. A2 and A3 check whether the closed-form estimators of Algorithm 2 recover the parameters that generated the graph. For each of 30 independent seeds at every target community size $S \in \{50, 100, 200, 500, 1000, 2000, 5000\}$, we draw per-community parameters at random – $m_i \in \{2, 4, 8, 16\}$, $\rho_i$ uniformly over $\{0.05, 0.10, \ldots, 0.95\}$, and $\sigma_i^2 = \phi_i m_i$ with $\phi_i \in \{1, 2, 4\}$ ($\phi_i = 1$ Poisson, $\phi_i > 1$ negative binomial) – generate a graph with $k = 12$ communities using CS, and re-estimate the parameters from it. Every graph is scored in two representations: *weighted* (multi-edges kept) and *simple* (duplicate edges pruned); the scatter uses the weighted representation, and the bias panel contrasts the two for communities with $N_i \geq 256$.

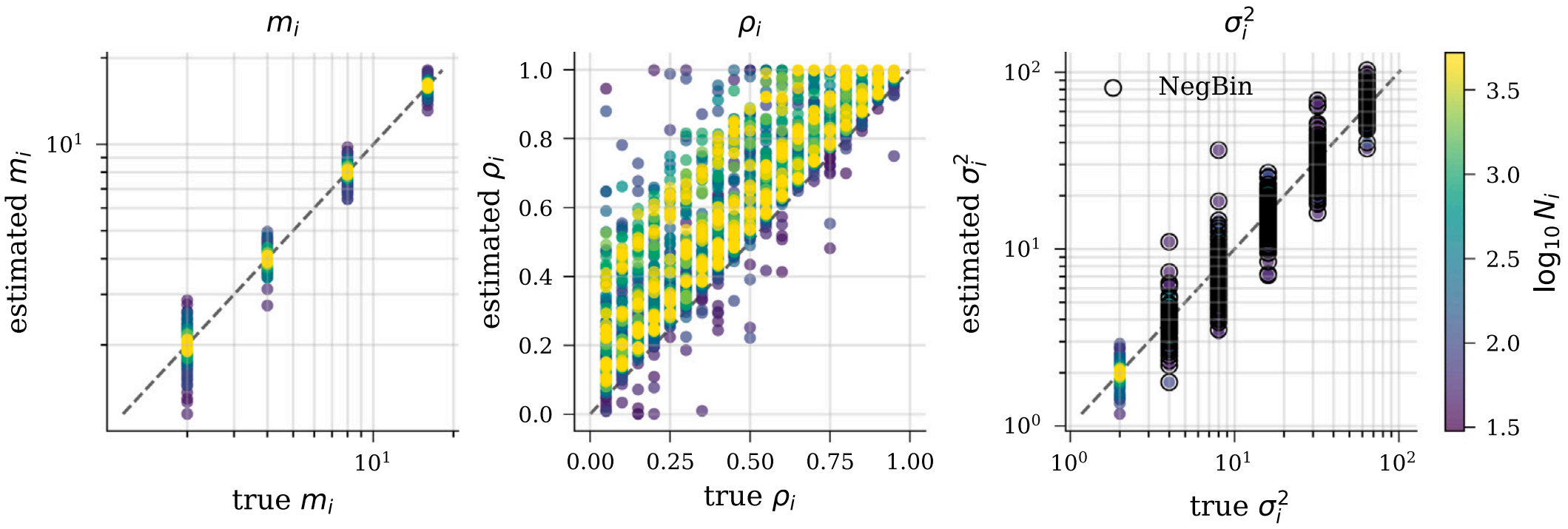

**Fig. A2.** Injected versus re-estimated citation Seeder parameters, one point per community, coloured by community size $N_i$. An open circle in the $\sigma_i^2$ panel marks a community generated by the negative-binomial branch ($\phi_i > 1$), as opposed to the Poisson fallback used at low over-dispersion. The dashed line is the identity.

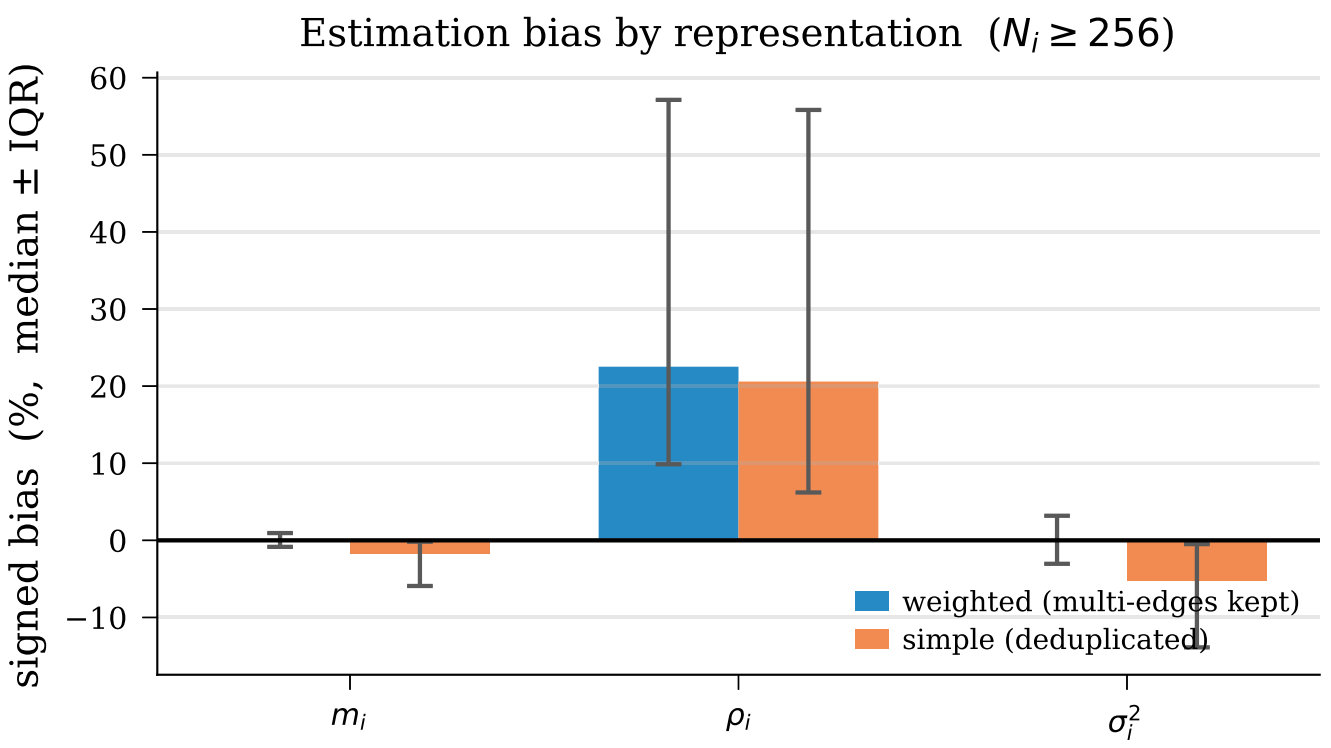

**Fig. A3.** Re-estimation bias by graph representation – weighted (multi-edges kept) versus simple (deduplicated edges) – for citation Seeder instances with identical input parameters.

The out-degree parameters $m_i$ and $\sigma_i^2$ are recovered tightly around the identity, with error shrinking steadily as community size grows; the preferentiality $\rho_i$ carries a systematic upward bias that plateaus rather than vanishing with size. This bias arises because the $\rho_i$ estimator equates the observed in-degree Gini coefficient with the value the model predicts analytically, so CS plants slightly higher effective preferentiality than the theoretical target. When comparing representations (Fig. A3), the weighted graph recovers $m_i$ and $\sigma_i^2$ with near-zero bias, while pruning to a simple graph trades a little $m_i/\sigma_i^2$ accuracy for a simple graph representation.

### A.3. Parameter effects

Fig. A4 verifies that each generative parameter controls the structural property it is meant to. Using the same baseline as Appendix A.1 ($k = 10$ communities of 1,000 nodes, 10 ensembles per point), we vary one parameter at a time – $m_i \in \{2, 4, 8, 16, 32\}$, $\rho_i \in \{0.1, 0.3, \dots, 0.9\}$, and $\phi_i \in \{1, 2, 4, 8\}$ – and plot each against its target statistic.

Every parameter moves its intended target monotonically and predictably: $\rho_i$ raises the in-degree Gini (citation inequality), $m_i$ sets the mean out-degree, and the over-dispersion ratio $\phi_i = \sigma_i^2/m_i$ controls the out-degree Fano factor. These clean one-to-one responses are what let the parameters serve as interpretable design knobs when a network is constructed from domain knowledge rather than fitted to data.

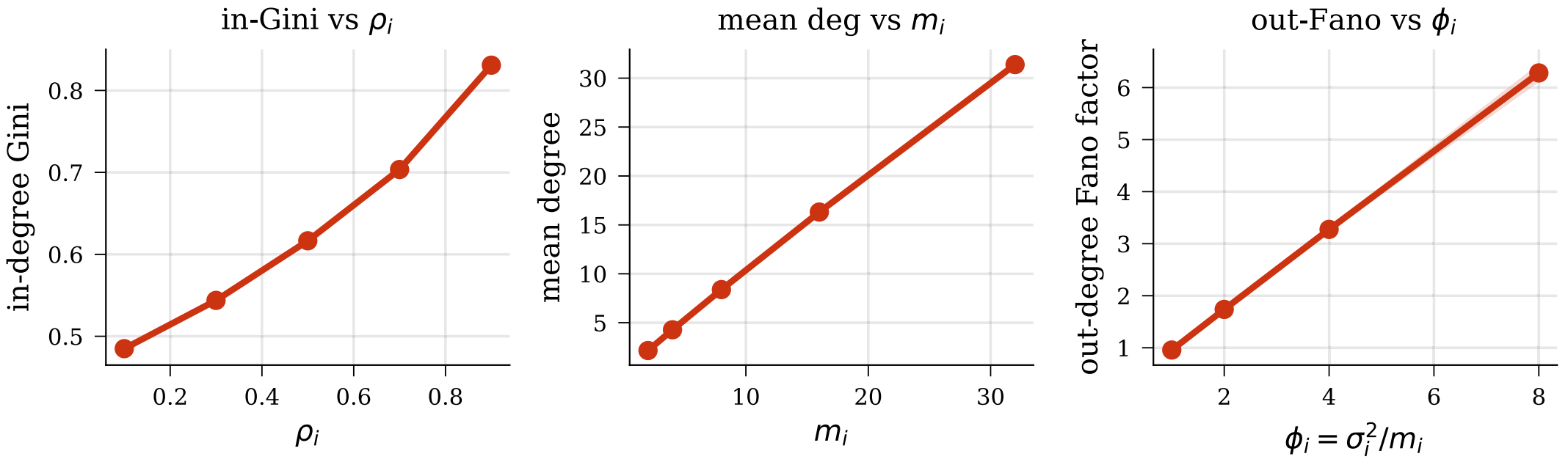

**Fig. A4.** Effect of each generative parameter on the structural property it targets, swept one at a time around the baseline operating point. Shaded bands span ±1 standard deviation over the ensemble, but they are effectively invisible.

### A.4. Forecasting fidelity

In this section, we compare CS with DC-SBM-nD and DC-SBM in the task of forecasting a network's growth from a temporal hold-out. We experiment on the OGBN-ArXiv network, as it is the cleaner of the two datasets equipped with real timestamps.

The protocol is as follows. For a training fraction $f \in \{0.5, 0.7, 0.9\}$, we take the early subgraph $G_{\text{early}}$ induced by the nodes whose timestamp falls at or below the $f$-quantile, and fit every method on $G_{\text{early}}$ alone. The only future quantity revealed to any method is the horizon size $N_{\text{late}}$. CS forecasts natively by running its growth process forward from $G_{\text{early}}$'s estimated parameters to $N_{\text{late}}$ and injecting back-edges at the early ratio $r_{\text{early}}$. DC-SBM is rescaled to $N_{\text{late}}$, preserving field shares, block affinities and mean degree, with DC-SBM-nD additionally cycle-broken at $r_{\text{early}}$; these are the two strongest baselines from the main evaluation. We score each forecast by the Wasserstein-1 distance between its in-degree distribution and that of the real future graph, averaged over 30 samples per method per horizon. This metric is $O(N)$, so it is computed on the full graphs without subsampling.

Fig. A5 reports forecast error against the training fraction. CS attains a lower in-degree Wasserstein-1 distance than both DC-SBM variants at every horizon, and its advantage widens as more history is observed, whereas the DC-SBM variants improve only marginally and plateau. This is the behaviour expected of a generative growth model as opposed to a descriptive one: additional early history sharpens the estimated growth parameters, whereas rescaling a fitted block model contributes little.

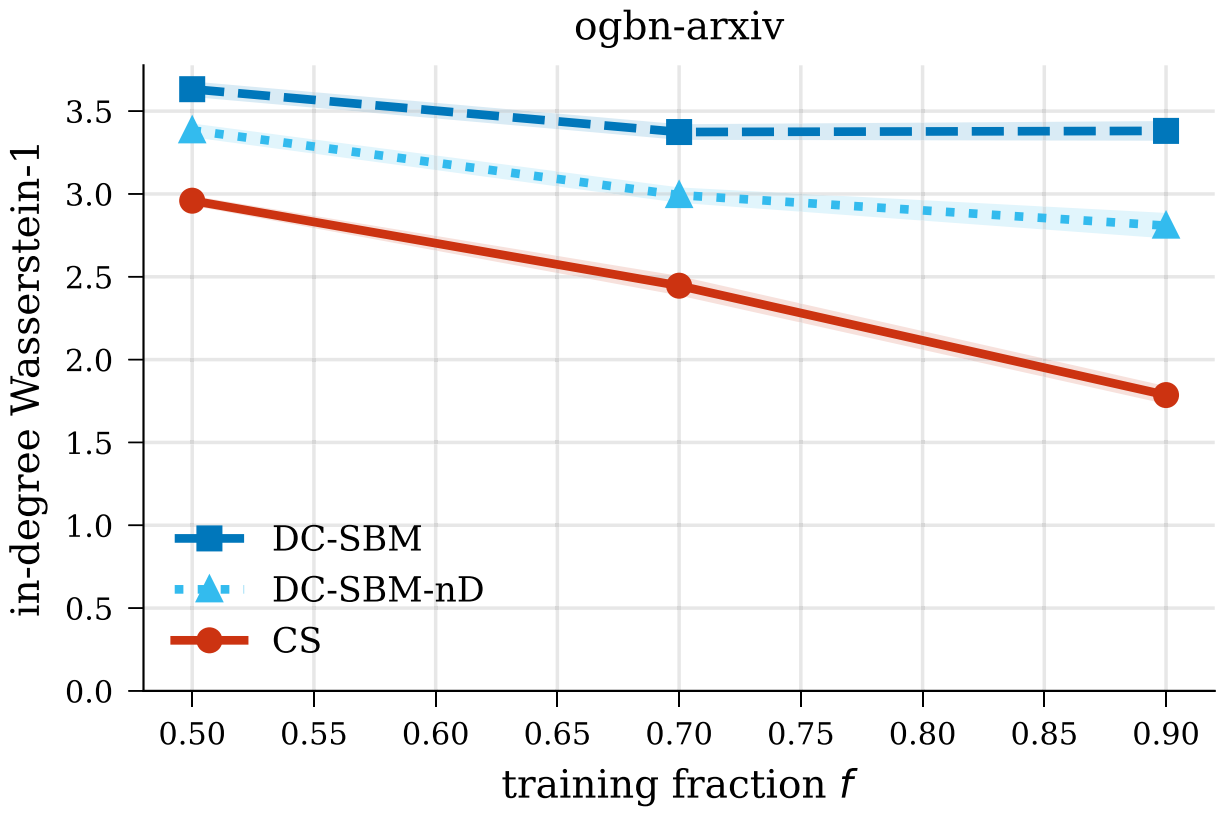

**Fig. A5.** Forecast error versus training fraction $f$ on OGBN-ArXiv. In-degree Wasserstein-1 distance to the real future graph (lower is better) for CS, DC-SBM and DC-SBM-nD; the CS band spans ±1 standard deviation over samples. CS improves as more history is observed, while the block models plateau.

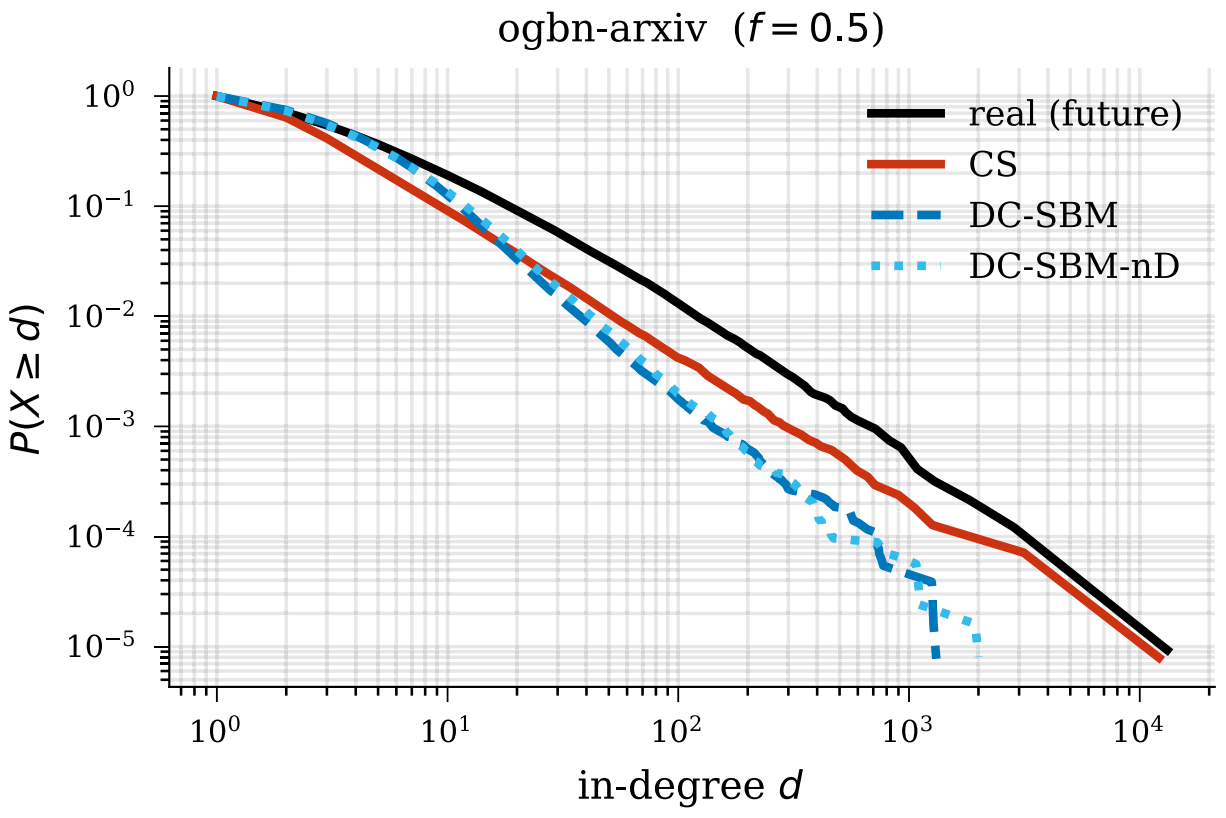

**Fig. A6.** In-degree complementary CDF at the hardest horizon ($f = 0.5$) on OGBN-ArXiv. CS reproduces the real future power-law tail; both DC-SBM variants underproduce high-in-degree nodes.

Fig. A6 shows the in-degree complementary CDF at the hardest horizon ($f = 0.5$), where only half the timeline is observed. CS tracks the real future power-law tail across its full range, while both DC-SBM variants fall well below it, underproducing high-in-degree hubs by an order of magnitude in tail probability.

We stress the scope of this result. It is a comparative validation on a single timestamped network, and it concerns the *degree distribution*: the experiment does not perform link prediction, and no claim of link-level forecasting is made. We also compare only against the strongest baselines from the main comparison, and we make no claim in terms of comparison with state-of-the-art forecasting models. Within that scope, CS extrapolates future degree growth more faithfully, supporting the interpretation of its parameters as generative rather than merely descriptive.

## Data availability

The authors are unable to or have chosen not to specify which data have been used.

## References

[1] L. Speidel, T. Takaguchi, N. Masuda, Community detection in directed acyclic graphs, Eur. Phys. J. B 88 (8) (2015) 203, https://doi.org/10.1140/epjb/e2015-60226-y
[2] D. Wang, C. Song, A.-L. Barabási, Quantifying long-term scientific impact, Science 342 (6154) (2013) 127–132, https://doi.org/10.1126/science.1237825
[3] M.E.J. Newman, The first-mover advantage in scientific publication, Europhysics Letters 86 (6) (2009) 68001, https://doi.org/10.1209/0295-5075/86/68001
[4] D. Price, Networks of scientific papers, Science 149 (3683) (1965) 510–515, https://doi.org/10.1126/science.149.3683.510
[5] R. Albert, A.-L. Barabási, Statistical mechanics of complex networks, Rev. Mod. Phys. 74 (1) (2002) 47–97, https://doi.org/10.1103/RevModPhys.74.47
[6] G. Bianconi, A.-L. Barabási, Competition and multiscaling in evolving networks, Europhysics Letters 54 (4) (2001) 436, https://doi.org/10.1209/epl/i2001-00260-6
[7] V. Nanumyan, C. Gote, F. Schweitzer, Multilayer network approach to modeling authorship influence on citation dynamics in physics journals, Phys. Rev. E 102 (3) (2020) 032303, https://doi.org/10.1103/physreve.102.032303
[8] F. Hu, L. Ma, X.-X. Zhan, Y. Zhou, C. Liu, H. Zhao, Z.-K. Zhang, The aging effect in evolving scientific citation networks, Scientometrics 126 (5) (2021) 4297–4309, https://doi.org/10.1007/s11192-021-03929-8
[9] N. Cai, L. Liu, J.-W. Cao, On collective dynamics of academic journals: polarization, fluctuation, and robustness, IEEE Trans. Comput. Soc. Syst. 12 (5) (2025) 3337–3348, https://doi.org/10.1109/TCSS.2025.3546474
[10] C. Lee, D. Wilkinson, A review of stochastic block models and extensions for graph clustering, Appl. Netw. Sci. 4 (1) (2019) 122, https://doi.org/10.1007/s41109-019-0232-2
[11] B. Karrer, M.E.J. Newman, Stochastic blockmodels and community structure in networks, Phys. Rev. E 83 (1) (2011) 016107, https://doi.org/10.1103/PhysRevE.83.016107
[12] B. Bollobás, A probabilistic proof of an asymptotic formula for the number of labelled regular graphs, Eur. J. Comb. 1 (4) (1980) 311–316, https://doi.org/10.1016/S0195-6698(80)80030-8
[13] A. Lancichinetti, S. Fortunato, F. Radicchi, Benchmark graphs for testing community detection algorithms, Phys. Rev. E 78 (4) (2008) 046110, https://doi.org/10.1103/physreve.78.046110
[14] J. Zhu, B. Chen, Y. Zeng, Community detection based on modularity and k-plexes, Inf. Sci. 513 (2020) 127–142, https://doi.org/10.1016/j.ins.2019.10.076
[15] J. Li, J. Yu, J. Li, H. Zhang, K. Zhao, Y. Rong, H. Cheng, J. Huang, Dirichlet graph variational autoencoder, Adv. Neural Inf. Process. Syst. 33 (2020) 5274–5283, https://dl.acm.org/doi/10.5555/3495724.3496167.
[16] J. Chang, D. Blei, Relational topic models for document networks, in: Proceedings of the Twelfth International Conference on Artificial Intelligence and Statistics (AISTATS'09), Vol. 5 of Proceedings of Machine Learning Research, 2009, pp. 81–88, http://proceedings.mlr.press/v5/chang09a.html.
[17] K. Nath, R. Shanmugam, V. Varadaranjan, ma-CODE: a multi-phase approach on community detection in evolving networks, Inf. Sci. 569 (2021) 326–343, https://doi.org/10.1016/j.ins.2021.02.068
[18] M. Drobyshevskiy, D. Turdakov, Random graph modeling: a survey of the concepts, ACM Comput. Surv. 52 (6) (2019) 131, https://doi.org/10.1145/3369782
[19] A. Bonifati, I. Holubová, A. Prat-Pèrez, S. Sakr, Graph generators, ACM Comput. Surv. 53 (2) (2020) 1–30, https://doi.org/10.1145/3379445
[20] S. Metzler, P. Miettinen, HyGen: generating random graphs with hyperbolic communities, Appl. Netw. Sci. 4 (1) (2019) 53, https://doi.org/10.1007/s41109-019-0166-8

[21] A. Carballo-Castro, M. Madeira, Y. Qin, D. Thanou, P. Frossard, Generating directed graphs with dual attention and asymmetric encoding, in: International Conference on Learning Representations (ICLR'26), 2026, pp. 134541–134589, https://proceedings.iclr.cc/paper_files/paper/2026/file/d99f9bd9c692e64c266acdc16b866e7f-Paper-Conference.pdf.
[22] A. Decelle, F. Krzakala, C. Moore, L. Zdeborová, Asymptotic analysis of the stochastic block model for modular networks and its algorithmic applications, Phys. Rev. E 84 (6) (2011) 066106, https://doi.org/10.1103/physreve.84.066106
[23] C. Moore, The computer science and physics of community detection: landscapes, phase transitions, and hardness, Bulletin of the EATCS 121 (2017), http://bulletin.eatcs.org/index.php/beatcs/article/view/480.
[24] E. Abbe, Community detection and stochastic block models: recent developments, J. Mach. Learn. Res. 18 (177) (2018) 1–86, http://jmlr.org/papers/v18/16-480.html.
[25] E. Abbe, C. Sandon, Achieving the KS threshold in the general stochastic block model with linearized acyclic belief propagation, Adv. Neural Inf. Process. Syst. 29 (2016) 1334–1342, https://dl.acm.org/doi/10.5555/3157096.3157246.
[26] F. Caltagirone, M. Lelarge, L. Miolane, Recovering asymmetric communities in the stochastic block model, IEEE Trans. Netw. Sci. Eng. 5 (3) (2017) 237–246, https://doi.org/10.1109/tnse.2017.2758201
[27] M. Wiliński, P. Mazzarisi, D. Tantari, F. Lillo, Detectability of macroscopic structures in directed asymmetric stochastic block model, Phys. Rev. E 99 (4) (2019) 042310, https://doi.org/10.1103/physreve.99.042310
[28] J. Sun, D. Ajwani, P.K. Nicholson, A. Sala, S. Parthasarathy, Breaking cycles in noisy hierarchies, in: Proceedings of the 2017 ACM on Web Science Conference (WebSci'17), Association for Computing Machinery, New York, NY, USA, 2017, pp. 151–160, https://doi.org/10.1145/3091478.3091495
[29] B. Karrer, M.E.J. Newman, Random graph models for directed acyclic networks, Phys. Rev. E 80 (4) (2009) 046110, https://doi.org/10.1103/PhysRevE.80.046110
[30] P. Sen, G.M. Namata, M. Bilgic, L. Getoor, B. Gallagher, T. Eliassi-Rad, Collective classification in network data, AI Mag. 29 (3) (2008) 93–106, https://doi.org/10.1609/aimag.v29i3.2157
[31] L. Brzozowski, M. Gagolewski, G. Siudem, B. Żogała-Siudem, The Price–Pareto growth model of networks with community structure, J. Informetr. 20 (3) (2026) 101850, https://doi.org/10.1016/j.joi.2026.101850
[32] M. Gagolewski, A framework for benchmarking clustering algorithms, Softwarex 20 (2022) 101270, https://doi.org/10.1016/j.softx.2022.101270
[33] M. Gagolewski, M. Bartoszuk, A. Cena, Are cluster validity measures (in)valid? Inf. Sci. 581 (2021) 620–636, https://doi.org/10.1016/j.ins.2021.10.004
[34] I. van Mechelen, A.-L. Boulesteix, R. Dangl, et al., A white paper on good research practices in benchmarking: the case of cluster analysis, Wiley Interdiscip. Rev. Data Min. Knowl. Discov. 13 (6) (2023) e1511, https://doi.org/10.1002/widm.1511
[35] T. Ullmann, C. Hennig, A.-L. Boulesteix, Validation of cluster analysis results on validation data: a systematic framework, Wiley Interdiscip. Rev. Data Min. Knowl. Discov. 12 (3) (2022) e1444, https://doi.org/10.1002/widm.1444
[36] A. Jaeger, D. Banks, Cluster analysis: a modern statistical review, Wiley Interdiscip. Rev. Comput. Stat. 15 (3) (2023) e1597, https://doi.org/10.1002/wics.1597
[37] H. Chen, Z. Yu, Q. Yang, J. Shao, Community detection in subspace of attribute, Inf. Sci. 602 (2022) 220–235, https://doi.org/10.1016/j.ins.2022.04.047
[38] L. Brisson, C. Bothorel, N. Duminy, DynBenchmark: customizable ground truths to benchmark community detection and tracking in temporal networks, in: Proceedings of France's International Conference on Complex Systems (FRCCS'25), 2025, pp. 74–85, https://doi.org/10.1007/978-3-032-00206-8_8
[39] G. Siudem, B. Żogała-Siudem, A. Cena, M. Gagolewski, Three dimensions of scientific impact, Proc. Natl. Acad. Sci. U. S. A. (PNAS) 117 (25) (2020) 13896–13900, https://doi.org/10.1073/pnas.2001064117
[40] L. Bertoli-Barsotti, M. Gagolewski, G. Siudem, B. Żogała-Siudem, Equivalence of inequality indices in the three-dimensional model of informetric impact, J. Informetr. 18 (4) (2024) 101566, https://doi.org/10.1016/j.joi.2024.101566
[41] P. Eades, X. Lin, W. Smyth, A fast and effective heuristic for the feedback arc set problem, Inf. Process. Lett. 47 (6) (1993) 319–323, https://doi.org/10.1016/0020-0190(93)90079-O
[42] J. Leskovec, A. Krevl, SNAP datasets: Stanford large network dataset collection, 2014, http://snap.stanford.edu/data.
[43] J. Tang, J. Zhang, L. Yao, J. Li, L. Zhang, Z. Su, Arnetminer: extraction and mining of academic social networks, in: Proceedings of the 14th ACM SIGKDD International Conference on Knowledge Discovery and Data Mining (KDD'08), Association for Computing Machinery, New York, NY, USA, 2008, pp. 990–998, https://doi.org/10.1145/1401890.1402008
[44] W. Hu, M. Fey, M. Zitnik, Y. Dong, H. Ren, B. Liu, M. Catasta, J. Leskovec, Open graph benchmark: datasets for machine learning on graphs, Adv. Neural Inf. Process. Syst. 33 (2020) 22118–22133, https://dl.acm.org/doi/10.5555/3495724.3497579.
[45] A. Bojchevski, S. Günnemann, Deep Gaussian embedding of graphs: unsupervised inductive learning via ranking, in: International Conference on Learning Representations (ICLR'18), 2018, https://iclr.cc/virtual/2018/poster/188.
[46] L. Brzozowski, UnPACD: unified patent and citation dataset, 2026, https://github.com/lukaszbrzozowski/unpacd.
[47] M.E.J. Newman, Networks, Oxford University Press, 2018, https://doi.org/10.1093/oso/9780198805090.001.0001
[48] P.W. Holland, K.B. Laskey, S. Leinhardt, Stochastic blockmodels: first steps, Soc. Netw. 5 (2) (1983) 109–137, https://doi.org/10.1016/0378-8733(83)90021-7
[49] S. Legramanti, T. Rigon, D. Durante, Bayesian testing for exogenous partition structures in stochastic block models, Sankhya A 84 (2022) 108–126, https://doi.org/10.1007/s13171-020-00231-2
[50] M. Rosvall, D. Axelsson, C.T. Bergstrom, The map equation, Eur. Phys. J. Spec. Top. 178 (1) (2009) 13–23, https://doi.org/10.1140/epjst/e2010-01179-1
[51] V. Traag, L. Waltman, N.J. van Eck, From louvain to Leiden: guaranteeing well-connected communities, Sci. Rep. 9 (2019) 5233, https://doi.org/10.1038/s41598-019-41695-z
[52] M. Friedman, The use of ranks to avoid the assumption of normality implicit in the analysis of variance, J. Am. Stat. Assoc. 32 (200) (1937) 675–701, https://doi.org/10.1080/01621459.1937.10503522